\documentclass[twocolumn,amsmath,amssymb,footinbib,superscriptaddress]{revtex4-2}\usepackage{lipsum} 
\usepackage{graphicx}
\usepackage{dcolumn}
\usepackage{bm}
\usepackage[colorlinks,linkcolor=blue,anchorcolor=blue,urlcolor=blue, citecolor=blue]{hyperref}
\usepackage{braket}
\usepackage{float}
\usepackage{xcolor}
\usepackage{etoolbox}

\newcommand{\affvqcc}{Vigo Quantum Communication Center, University of Vigo, Vigo E-36310, Spain}
\newcommand{\affuvigo}{Escuela de Ingeniería de Telecomunicación, Department of Signal Theory and Communications, University of Vigo, Vigo E-36310, Spain}
\newcommand{\affatlantic}{atlanTTic Research Center, University of Vigo, Vigo E-36310, Spain}
\newcommand{\afftoyama}{Faculty of Engineering, University of Toyama, Gofuku 3190, Toyama 930-8555, Japan}

\begin{document}
	\title{Secret key rate bounds for quantum key distribution with non-uniform phase randomization}

	\author{Xoel Sixto}
        \email{xsixto@com.uvigo.es}
	\affiliation{\affvqcc} \affiliation{\affuvigo} \affiliation{\affatlantic} 
	\author{Guillermo Currás-Lorenzo}
	 	\affiliation{\afftoyama}
        \author{Kiyoshi Tamaki}
        \affiliation{\afftoyama} 
        \author{Marcos Curty}
        \affiliation{\affvqcc} \affiliation{\affuvigo} \affiliation{\affatlantic} 

\date{\today}

\begin{abstract}
Decoy-state quantum key distribution (QKD) is undoubtedly the most efficient solution to handle multi-photon signals emitted by laser sources, and provides the same secret key rate scaling as ideal single-photon sources. It requires, however, that the phase of each emitted pulse is uniformly random. This might be difficult to guarantee in practice, due to inevitable device imperfections and/or the use of an external phase modulator for phase randomization, which limits the possible selected phases to a finite set. Here, we investigate the security of decoy-state QKD with arbitrary, continuous or discrete, non-uniform phase randomization, and show that this technique is quite robust to deviations from the ideal uniformly random scenario. For this, we combine a novel parameter estimation technique based on semi-definite programming, with the use of basis mismatched events, to tightly estimate the parameters that determine the achievable secret key rate. In doing so, we demonstrate that our analysis can significantly outperform previous results that address more restricted scenarios. 
\end{abstract}

\maketitle

\section{Introduction}\label{Introduction}

Quantum key distribution (QKD) is a method for securely establishing symmetric cryptographic keys between two distant parties (so-called Alice and Bob)~\cite{extra1,extra2,extra3}. Its security is based on principles of quantum mechanics, such as the no-cloning theorem~\cite{cloning}, which guarantee that any attempt by an eavesdropper (Eve) to learn information about the distributed key inevitably introduces detectable errors. Importantly, when combined with the one-time-pad encryption scheme~\cite{Vernam}, QKD provides information-theoretically secure communications. 

The field of QKD has made much progress in recent years, both theoretically and experimentally, leading to the first deployments of  metropolitan and intercity QKD networks~\cite{nueva1, nueva2, nueva3, nueva4}. Despite these remarkable achievements, there are still certain challenges that need to be overcome for the widespread adoption of this technology. One of these challenges is to close the existing security gap between theory and practice. This is so because QKD security proofs, typically consider assumptions that the actual experimental implementations do not satisfy. Such discrepancies could create security loopholes or so-called side channels, which might be exploited by Eve to compromise the security of the generated key without being detected.

Indeed, practical QKD transmitters usually emit phase-randomized weak coherent pulses (PR-WCPs) generated by laser sources. These pulses might contain more than one photon prepared in the same quantum state. In this scenario, Eve is no longer limited by the no-cloning theorem, because multi-photon signals provide her with perfect copies of the signal photon. As a result, it can be shown that the secret key rate of the BB84 protocol~\cite{BB84} with PR-WCPs scales quadratically with the system's transmittance due to the photon-number-splitting (PNS) attack~\cite{PNS1,PNS2}. This attack provides Eve with full information about the part of the key generated with the multi-photon pulses, without introducing any error. 

To overcome this limitation, the most efficient solution today is undoubtedly the decoy-state method~\cite{decoy1,decoy2,decoy3}, in which Alice varies at random the intensity of the PR-WCPs that she sends to Bob. This allows them to better estimate the behavior of the quantum channel. Indeed, using the observed measurement statistics associated to different intensity settings, Alice and Bob can tightly estimate the yield and phase error rate of the single-photon pulses, from which the secret key is actually distilled. As a result, the decoy-state method delivers a secret key rate that scales linearly with the channel transmittance~\cite{decoy1,decoy2,decoy3,linear}, matching the scaling achievable with ideal single-photon sources. This technique has been extensively demonstrated in multiple recent experiments~\cite{demos1,demos2,demos3,demos4,demos5,demos6,gain_switch_5}, including satellite links \cite{sat1,sat2} and the use of photonic integrated circuits~\cite{chip1,chip2,chip3,chip4}. Also, decoy-state QKD setups are currently offered commercially by several companies~\cite{IDQ, Toshiba, QuantumCTek, ThinkQ, QTI}, which highlights its importance.

Importantly, phase randomization means that the phase, $\theta$, of each generated WCP must be uniformly random in $[0,2\pi)$. That is, its probability density function (PDF), $g(\theta)$, must satisfy $g(\theta)=1/2\pi$. However, none of the two main methods used today to generate PR-WCPs fulfill this condition exactly. Precisely, in those configurations that drive the laser source under gain-switching conditions~\cite{gain_switch_1,gain_switch_2,gain_switch_3,gain_switch_4,gain_switch_5, Valivarthi2017, Yuan2018}, device imperfections can prevent the phases $\theta$ from being uniformly distributed. Similarly, in those configurations that use an external phase modulator for this purpose~\cite{generation1, generation2,Sibson2017,PhysRevX.8.021009}, only a discrete number of phases is selected. Both scenarios violate a crucial assumption of the decoy-state technique. 

The discrete phase-randomization scenario has been analyzed in~\cite{Lo_Ma} (see also~\cite{Guillermo_2021}). This work assumes evenly distributed discrete random phases in $[0,2\pi)$, {\it i.e.}, it considers that $g(\theta)$ satisfies
\begin{equation}
\label{eq:dis_perfect}
g(\theta) = \frac{1}{N} \sum_{k=0}^{N-1} \delta(\theta-\theta_{k}),
\end{equation}
where $\delta(x)$ represents the Dirac delta function, and $\theta_{k}=2 \pi k/N$, with $N$ being the total number of selected phases. Under this assumption,~\cite{Lo_Ma} shows that it is possible to approximate the secret key rate achievable in the ideal situation where $g(\theta)=1/2\pi$, with around $N=10$ random phases. While this result is remarkable, in practice, inevitable imperfections of the phase modulator and electronic noise might prevent the phases $\theta$ from being {\it exactly} evenly distributed, thus invalidating the application of the results presented in~\cite{Lo_Ma} to a real setup.

In this paper, we consider the more realistic and practical scenario in which $g(\theta)$ could be an arbitrary, continuous or discrete, PDF, due to imperfections in the phase-randomization process, and we provide asymptotic secret key rates for this general situation. In our derivations, for simplicity, we consider collective attacks, but our results are also valid against coherent attacks due to the quantum de Finetti theorem~\cite{finetti}. The key ingredients of our study are two: the use of basis mismatched events ({\it i.e.}, events in which Alice and Bob select different bases), and a novel parameter estimation technique based on semi-definite programming (SDP), very recently introduced in~\cite{Guillermo}. Importantly, we show that the combination of these two ingredients permits a tight estimation of the relevant parameters needed to evaluate the secret key rate in the scenario considered here. In doing so, we find that the decoy-state method is indeed very robust to imperfect phase randomization even with an arbitrary, continuous or discrete, $g(\theta)$. Remarkably, for the ideal discrete phase-randomization case described by Eq.~(\ref{eq:dis_perfect}), our analysis delivers significantly higher secret key rates than those provided by the seminal analysis presented in~\cite{Lo_Ma}. Or, to put it in other words, it requires fewer random bits for phase selection to achieve a similar performance.

The paper is organized as follows. In Sec.~\ref{Generated}, we describe the quantum states emitted by Alice when $\theta$ follows an arbitrary PDF, $g(\theta)$. Then, in Sec.~\ref{Protocol_Key} we introduce the decoy-state protocol considered, together with its asymptotic secret key rate formula. Next, in Sec.~\ref{Parameter}, we present the parameter estimation technique based on SDP, as well as on the use of basis mismatched events, to calculate the different parameters required to evaluate the secret key rate. Then, in Sec.~\ref{numerical} we simulate the achievable secret key rate for various functions $g(\theta)$ of practical interest, both for the cases in which this function is fully (or only partially) characterized. Sec.~\ref{conc} concludes the paper with a summary. The paper includes as well some Appendixes with additional calculations.

\section{Phase randomization with an arbitrary $g(\theta)$}\label{Generated}

In this section, we describe the quantum states emitted by Alice when each of them has a phase $\theta$ that follows an arbitrary PDF, $g(\theta)$. 

In particular, a WCP of intensity $\mu$ and phase $\theta$ can be written in terms of the Fock basis as
\begin{equation}
\label{eq:coherent}
|\sqrt{\mu}e^{i \theta}\rangle=e^{-\frac{\mu}{2}} \sum_{n=0}^{\infty}\frac{\left(\sqrt{\mu}e^{i \theta}\right)^n}{\sqrt{n !}}|n\rangle,
\end{equation}
where $|n\rangle$ represents a Fock state with $n$ photons. 

If Alice selects the phase $\theta$ of each generated signal independently and at random according to $g(\theta)$, its state is simply given by
\begin{equation}
\label{eq:general_state}
\rho^{\mu}_{[g(\theta)]}= \int_0^{2 \pi} g(\theta){\hat P}(|\sqrt{\mu} e^{i \theta}\rangle)d \theta,
\end{equation}
with ${\hat P}(|\phi\rangle)=|\phi\rangle\langle\phi|$. 

Any quantum state can always be diagonalised in a certain orthonormal basis. For the states given by Eq.~(\ref{eq:general_state}), we shall denote the elements of such basis by $|\psi_{n, \mu, g(\theta)}\rangle$, since, in general, they might depend on both the intensity $\mu$ and the function $g(\theta)$. Here, the subscript $n$ simply identifies the different elements of the basis, which are not necessarily the Fock states. This means, in particular, that we can rewrite the states given by Eq.~(\ref{eq:general_state}) as follows
\begin{equation}
\label{eq:diagonal}
\rho^\mu_{[g(\theta)]}=\sum_{n=0}^{\infty} p_{n| \mu, g(\theta)}{\hat P}(|\psi_{n, \mu, g(\theta)}\rangle),
\end{equation}
where the coefficients $p_{n| \mu, g(\theta)}\geq{}0$ satisfy $\sum_{n=0}^{\infty} p_{n| \mu, g(\theta)}=1$. That is, these coefficients can be interpreted as the probability with which, in a certain time instance, Alice emits the state $|\psi_{n, \mu, g(\theta)}\rangle$, given that she chose the intensity $\mu$ and $\theta$ follows the PDF $g(\theta)$. 

For instance, in the ideal scenario where $g(\theta)$ is uniformly random in $[0,2\pi)$, the emitted signals are a Poisson mixture of Fock states given by
\begin{eqnarray}
\label{eq:general_state_ideal}
\rho^{\mu}_{[\frac{1}{2\pi}]}&=& \frac{1}{2\pi}\int_0^{2 \pi}{\hat P}(|\sqrt{\mu} e^{i \theta}\rangle)d \theta \nonumber \\
&=&e^{-\mu}\sum_{n=0}^\infty \frac{\mu^n}{n !} {\hat P}(|n\rangle),
\end{eqnarray}
{\it i.e.} $p_{n| \mu, 1/2\pi}=e^{-\mu}\mu^n/(n!)$ and $|\psi_{n, \mu, 1/2\pi}\rangle=|n\rangle$.

\section{Protocol description and key generation rate}\label{Protocol_Key}

For concreteness, we shall assume that Alice and Bob implement a decoy-state BB84 scheme with three different intensity settings $\{s, \nu, \omega \}$ in each basis, with $s>\nu>\omega\geq{}0$. Moreover, we consider that they generate secret key only from those events in which both of them select the $Z$ basis and Alice chooses the signal intensity setting $s$. This is the most typical configuration of the decoy-state BB84 protocol. We remark, however, that the analysis below could be straightforwardly adapted to other protocol configurations, or to other combinations of intensity settings. 

In each round of the protocol, Alice probabilistically chooses a bit value $b\in\{0,1\}$ with probability $p_b=1/2$, a basis $\alpha\in\{Z,X\}$ with probability $p_\alpha$, an intensity value $\mu \in \{s, \nu, \omega \}$ with probability $p_{\mu}$, and a random phase $\theta$ according to the PDF given by $g(\theta)$. Then, she generates a WCP of intensity $\mu$ and phase $\theta$, $|\sqrt{\mu}e^{i \theta}\rangle$, and applies an operation that encodes her bit and basis choices $b$ and $\alpha$ into the pulse. From Eve's perspective, these states are described by Eq.~(\ref{eq:diagonal}) due to her ignorance about the selected phase $\theta$. On the receiving side, Bob measures each arriving signal using a basis $\alpha\in\{Z,X\}$, which he selects with probability $p_\alpha$. We shall assume the basis independent detection efficiency condition throughout the paper. That is, the probability that Bob obtains a conclusive measurement outcome does not depend on his basis choice.

Once the quantum communication phase of the protocol ends, Alice and Bob broadcast (via an authenticated classical channel) both the intensity and basis settings selected for each detected signal. The results related to those detected signals in which both of them used the $Z$ basis with intensity setting $s$ constitute the sifted key. For the detected rounds in which Bob chose the $X$ basis, Alice reveals her bit values $b$ and Bob announces his corresponding measurement outcomes. This data is used for parameter estimation, {\it i.e.}, to determine the relevant quantities needed to evaluate the secret key rate formula. Finally, Alice and Bob apply error correction and privacy amplification to the sifted key to obtain a final secret key, following the standard post-processing procedure in QKD~\cite{extra1,extra2,extra3}. For a more detailed description of the protocol steps of a decoy-state BB84 scheme, we refer the reader to {\it e.g.}~\cite{linear}.

In the ideal scenario where $g(\theta)=1/2\pi$, Alice's state preparation process is equivalent to emitting Fock states $|n\rangle$ with a Poisson distribution of mean equal to the intensity setting $\mu$ selected, as shown by Eq.~(\ref{eq:general_state_ideal}). In this situation, both the single-photon and vacuum pulses with the intensity setting $s$ contribute to secret bits~\cite{vacuum}. The multi-photon signals are insecure due to the PNS attack. Similarly, when $\theta$ follows an arbitrary PDF, $g(\theta)$, and Alice chooses the intensity setting $\mu$, from Eq.~(\ref{eq:diagonal}) we have that her state preparation process is equivalent to generating pure states $|\psi_{n, \mu, g(\theta)}\rangle$ with probability $p_{n| \mu, g(\theta)}$. The closer the function $g(\theta)$ is to a uniform distribution, the closer the signals (probabilities) $|\psi_{n, \mu, g(\theta)}\rangle$ ($p_{n| \mu, g(\theta)}$) are to the Fock states $|n\rangle$ (probabilities $e^{-\mu}\mu^n/n!$). In this scenario, Alice and Bob can in principle distill secret bits from any $|\psi_{n, \mu, g(\theta)}\rangle$ with $\mu=s$, though the main contribution would mainly arise from those with indexes $n=0,1$, which are the ones closer to vacuum and single-photon pulses. These are the contributions that we consider below. Indeed, for the examples studied in Sec.~\ref{numerical}, we have observed that the secret key rate improvement that one might obtain when considering $n>1$ is essentially negligible.

This means, in particular, that the asymptotic secret key rate formula for the decoy-state BB84 protocol considered can be written as~\cite{decoy3,GLLP,vacuum}
\begin{eqnarray}
\label{eq:skr_general}
R&\geq& p_Z^2p_s\Bigg\{\sum_{n=0}^{\infty} p_{n| s, g(\theta)}Y_{n, s, g(\theta)}^{Z}\left[1-h\left(e_{n, s, g(\theta)}\right)\right] \nonumber \\
&-&f Q_{s, g(\theta)}^Z h\left(E_{s, g(\theta)}^Z\right)\Bigg\} \nonumber \\
&\geq& p_Z^2p_s\Bigg\{\sum_{n=0}^{1} p_{n| s, g(\theta)}^{\text{L}}Y_{n, s, g(\theta)}^{Z, \text{L}}\left[1-h\left(e_{n, s, g(\theta)}^{\mathrm{U}}\right)\right] \nonumber \\
&-&f Q_{s, g(\theta)}^Z h\left(E_{s, g(\theta)}^Z\right)\Bigg\},
\end{eqnarray}
where $Y_{n, s, g(\theta)}^{Z}$ denotes the yield associated to the state $|\psi_{n, s, g(\theta)}\rangle$ encoded (and measured) in the $Z$ basis, {\it i.e.}, the probability that Bob observes a detection click in his measurement apparatus conditioned on Alice and Bob selecting the $Z$ basis and Alice preparing the state $|\psi_{n, s, g(\theta)}\rangle$; the parameter $e_{n, s, g(\theta)}$ represents the phase error rate of these latter signals; $h(x)=-x\log_2{(x)}-(1-x)\log_2{(1-x)}$ is the binary Shannon entropy function; the quantity $f$ is the efficiency of the error correction protocol; $Q_{s, g(\theta)}^Z$ is the overall gain of the signals emitted conditioned on Alice selecting the intensity $s$ and Alice and Bob choosing the $Z$ basis, {\it i.e.}, the probability that Bob observes a detection click conditioned on Alice sending him such signals; and $E_{s, g(\theta)}^Z$ is the overall quantum bit error rate (QBER) associated to these latter signals. Moreover, in Eq.~(\ref{eq:skr_general}), the superscript L (U) refers to a (an) lower (upper) bound. 

The quantities $Q_{s, g(\theta)}^Z$ and $E_{s, g(\theta)}^Z$ are directly observed in the experiment. In principle, the probabilities $p_{n| s, g(\theta)}$ could also be known, and depend on the state preparation process. However, in practice it might be difficult to find their value analytically. Instead, in the next section we present a simple method to obtain a lower bound, $p_{n| s, g(\theta)}^{\mathrm{L}}$, on these quantities. There, we also explain how to estimate the parameters $Y_{n, s, g(\theta)}^{Z, \mathrm{L}}$ and $e_{n, s, g(\theta)}^{\mathrm{U}}$, with $n=0,1$, which are needed to evaluate Eq.~(\ref{eq:skr_general}). 

\section{Parameter estimation}\label{Parameter}

The analysis follows the techniques very recently introduced in~\cite{Guillermo} in the context of phase correlations in gain-switched lasers. For simplicity, below we introduce the main results and refer the reader to Appendixes~\ref{app:infinite} and~\ref{app:bounds_SDP} for the detailed derivations.

\subsection{Lower bound on the yields $Y_{n, s, g(\theta)}^{Z}$}

In Appendix~\ref{app:infinite} it is shown that a lower bound on the yields $Y_{n, s, g(\theta)}^{Z}$ can be obtained by solving the following SDP:
\begin{equation}
\label{eq:SDP_yield_inf}
\begin{array}{cl}
\min_ {J_{Z}} & \operatorname{Tr}\left[{\hat P}(|\psi_{n, s, g(\theta)}\rangle)J_{Z}\right] \\
\text { s.t. } & \operatorname{Tr}\left[\rho^{\mu}_{[g(\theta)]} J_{Z}\right]=Q_{\mu, g(\theta)}^Z, \quad \forall \mu\in\{s, \nu, \omega\} \\
& 0 \leq J_{Z} \leq \mathbb{I}.
\end{array}
\end{equation}
The states $|\psi_{n, s, g(\theta)}\rangle$ and $\rho^\mu_{[g(\theta)]}$ are known in principle but inaccessible and depend on the intensity setting selected by Alice and on the function $g(\theta)$. Also, as already mentioned, the gains $Q_{\mu, g(\theta)}^Z$ are directly observed experimentally in a realization of the protocol. That is, the only unknown in Eq.~(\ref{eq:SDP_yield_inf}) is the positive semi-definite operator $J_{Z}$ over which the minimization takes place. Let $J_{Z}^*$ denote the solution to the SDP given by Eq.~(\ref{eq:SDP_yield_inf}). Then, we find that
\begin{equation}\label{sat_morn1}
Y_{n, s, g(\theta)}^{Z}\geq\operatorname{Tr}\left[{\hat P}(|\psi_{n, s, g(\theta)}\rangle)J_{Z}^*\right]:= Y_{n, s, g(\theta)}^{Z, {\rm L}}.
\end{equation}

\subsection{Upper bound on the phase-error rates $e_{n, s, g(\theta)}$}

The phase-error rates, $e_{n, s, g(\theta)}$, are defined by means of a virtual protocol~\cite{Koashi2009}. For this, we shall consider the standard assumption in which the efficiency of Bob’s measurement is independent of his basis choice. Then, for those rounds in which both Alice and Bob select the $Z$ basis and Alice generates the $n$-th eigenstate $|\psi_{n, s, g(\theta)}\rangle$, we can equivalently describe her state preparation process as follows. First, she prepares the following bipartite entangled state
\begin{equation}
\label{purif}
|\Psi^Z_{n, s, g(\theta)}\rangle=\frac{1}{\sqrt{2}}\left(|0_Z\rangle_A \hat{V}_{0_Z}+|1_Z\rangle_A \hat{V}_{1_Z}\right)|\psi_{n, s, g(\theta)}\rangle, 
\end{equation}
where $\hat{V}_{b_\alpha}$, with $b=0,1$ and $\alpha\in\{Z,X\}$, denotes the encoding operation corresponding to the $\alpha$ basis and the bit value $b$. Although our analysis is valid for any $\{\hat{V}_{b_{\alpha}}\}$, for simplicity, in our simulations, we assume that these operators, are ideal BB84 encoding operators, given by $\hat{V}_{0_Z}|n\rangle=|n\rangle|0\rangle, \hat{V}_{1_Z}|n\rangle=|0\rangle|n\rangle,$
\begin{equation}
\begin{gathered}
\hat{V}_{0_X}|n\rangle=\sum_k \frac{1}{\sqrt{2^n}} \sqrt{\left(\begin{array}{c}
n \\
k
\end{array}\right)}|k\rangle|n-k\rangle, \\
\hat{V}_{1_X}|n\rangle=\sum_k(-1)^k \frac{1}{\sqrt{2^n}} \sqrt{\left(\begin{array}{l}
n \\
k
\end{array}\right)}|k\rangle|n-k\rangle .
\end{gathered}
\end{equation}
Next, she measures her ancilla system $A$ in Eq.~(\ref{purif}) in the orthonormal basis $\{|0_Z\rangle, |1_Z\rangle\}$ to learn the bit value encoded, and sends the other system to Bob, who measures it in the $Z$ basis. 

In this situation, the phase-error rate $e_{n, s, g(\theta)}$ corresponds to the bit error rate that Alice and Bob would observe if Alice (Bob)  instead performed an $X$ basis measurement on the ancilla system $A$ (arriving signal). If Alice performs a $X$ basis measurement on her system $A$, this is equivalent to emitting the states
\begin{eqnarray}
\label{eq:unnormalized}
|\lambda^{\mathrm{virtual}}_{ \Delta, n, s, g(\theta)}\rangle &\propto&|\bar{\lambda}^{\mathrm{virtual}}_{ \Delta, n, s, g(\theta)}\rangle={ }_A\langle\Delta_X |\Psi^Z_{n, s, g(\theta)}\rangle \nonumber\\
&=&\frac{1}{2}\left[\hat{V}_{0_Z}+(-1)^\Delta \hat{V}_{1_Z}\right]|\psi_{n, s, g(\theta)}\rangle,
\end{eqnarray}
with probability $p^{\mathrm{virtual}}_{\Delta, n, s, g(\theta)}=\||\bar{\lambda}^{\mathrm{virtual}}_{\Delta, n, s, g(\theta)}\rangle \|^2$, where $\Delta \in\{0,1\}$ and $|\Delta_X\rangle=\left[|0_Z\rangle+(-1)^\Delta|1_Z\rangle\right] / \sqrt{2}$. 
Let $Y_{\Delta, n, s, g(\theta)}^{ (\Delta \oplus 1)_X, \mathrm{virtual}}$ denote the probability that Bob obtains the measurement outcome $(\Delta \oplus 1)_X$ when he performs an $X$ basis measurement on the arriving signal conditioned on Alice emitting the state $|\lambda^{\mathrm{virtual}}_{ \Delta, n, s, g(\theta)}\rangle$. That is, this event corresponds to a phase error. Then, the phase error rate $e_{n, s, g(\theta)}$ can be written as
\begin{equation}
\label{eq:phase_virtual}
e_{n, s, g(\theta)}=\frac{1}{Y_{n, s, g(\theta)}^{Z}}\sum_{\Delta=0}^1 p^{\mathrm{virtual}}_{\Delta, n, s, g(\theta)}Y_{\Delta, n, s, g(\theta)}^{ (\Delta \oplus 1)_X, \mathrm{virtual}}.
\end{equation}

In Appendix~\ref{app:infinite}, it is shown that an upper bound on the quantity $p^{\mathrm{virtual}}_{\Delta, n, s, g(\theta)}Y_{\Delta, n, s, g(\theta)}^{ (\Delta \oplus 1)_X, \mathrm{virtual}}$ can be obtained by solving the following SDP:
\begin{equation}
\label{eq:SDP_error_inf}
\begin{aligned}
\max _{L_{(\Delta \oplus 1)_X}} & \operatorname{Tr}\left[{\hat P}(|\bar{\lambda}^{\text {virtual}}_{\Delta, n, s, g(\theta)}\rangle) L_{(\Delta \oplus 1)_X}\right] \\
\text { s.t. } & \operatorname{Tr}\left[\hat{V}_{b_{\alpha}} \rho^\mu_{[g(\theta)]} \hat{V}_{b_{\alpha}}^{\dagger} L_{(\Delta \oplus 1)_X}\right]=Q_{\mu, g(\theta), b_{\alpha}}^{(\Delta \oplus 1)_X}, \\
&\forall \mu \in \{s, \nu, \omega \}, \forall b\in\{0,1\}, \forall \alpha\in\{Z,X\} \\
& 0 \leq L_{(\Delta \oplus 1)_X} \leq \mathbb{I},
\end{aligned}
\end{equation}
where $\rho^\mu_{[g(\theta)]}$ is given by Eq.~\eqref{eq:diagonal}, and $Q_{\mu, g(\theta), b_{\alpha}}^{(\Delta \oplus 1)_X}$ denotes the probability that Bob observes the result $(\Delta \oplus 1)_X$ with his $X$ basis measurement given that Alice chose the intensity setting $\mu$, the basis $\alpha$, the bit value $b$, and the phases $\theta$ follow the PDF $g(\theta)$. We note that Eq.~(\ref{eq:SDP_error_inf}) includes constraints provided by basis mismatched events~\cite{Tamaki} in which Alice prepares the signals in the $Z$ basis and Bob measures them in the $X$ basis, which may result in a tighter estimation. This is because, in general, $|\lambda_{\Delta, n, s, g(\theta)}^{\text {virtual}}\rangle \neq \hat{V}_{\Delta_X}|\psi_{n, s, g(\theta)}\rangle$, and ${\hat P}(|\lambda^{\text{virtual}}_{\Delta, n,s, g(\theta)})$ may be better approximated by an operator-form linear combination of both $Z$-encoded and $X$-encoded states, rather than just the latter. 

Importantly, the states $|\bar{\lambda}^{\text {virtual}}_{\Delta, n, s, g(\theta)}\rangle$ and $\rho^\mu_{[g(\theta)]}$, as well as the operators $\hat{V}_{b_{\alpha}}$, are known and depend on Alice's state preparation process. The gains $Q_{\mu, g(\theta), b_{\alpha}}^{(\Delta \oplus 1)_X}$ are directly observed in a realization of the protocol. That is, the only unknown in Eq.~(\ref{eq:SDP_error_inf}) is the positive semi-definite operator $L$ over which the maximization takes place. 

Let $L_{(\Delta \oplus 1)_X}^{*}$ denote the solution to the SDP given by Eq.~(\ref{eq:SDP_error_inf}). Then, we have that
\begin{equation}
p^{\mathrm{virtual}}_{\Delta, n, s, g(\theta)}Y_{\Delta, n, s, g(\theta)}^{ (\Delta \oplus 1)_X, \mathrm{virtual}}\leq\operatorname{Tr}\left[{\hat P}(|\bar{\lambda}^{\text {virtual}}_{\Delta, n, s, g(\theta)}\rangle) L_{(\Delta \oplus 1)_X}^{*}\right].
\end{equation}
That is, 
\begin{eqnarray}
e_{n, s, g(\theta)}&\leq&\frac{1}{Y_{n, s, g(\theta)}^{Z, {\rm L}}}\sum_{\Delta=0}^1 \operatorname{Tr}\left[{\hat P}(|\bar{\lambda}^{\text {virtual}}_{\Delta, n, s, g(\theta)}\rangle) L_{(\Delta \oplus 1)_X}^{*}\right] \nonumber \\
&:=&e_{n, s, g(\theta)}^{\mathrm{U}}. 
\end{eqnarray}

\subsection{Solving Eqs.~(\ref{eq:SDP_yield_inf})-(\ref{eq:SDP_error_inf}) numerically}\label{sub_projection}

Solving numerically the SDPs presented above is difficult for two main reasons. Firstly, they are infinitely dimensional, because the states $\rho^\mu_{[g(\theta)]}$ are infinite-dimensional. Secondly, this also renders the calculation of the eigendecomposition of $\rho^\mu_{[g(\theta)]}$ given by Eq.~(\ref{eq:diagonal}) a difficult task. To overcome these two limitations, we follow a technique recently introduced in~\cite{Guillermo, Nahar} (see also \cite{proj}), which consists in projecting the states $\rho^\mu_{[g(\theta)]}$ onto a finite-dimensional subspace that contains up to $M$ photons. We shall denote the projected states as
\begin{equation}
\label{eq:discrete_proy}
\rho^\mu_{[g(\theta)], M}=\frac{\Pi_M \rho^\mu_{[g(\theta)]} \Pi_M}{\operatorname{Tr}\left[\Pi_M \rho^\mu_{[g(\theta)]} \Pi_M\right]},
\end{equation}
where $\Pi_M=\sum_{n=0}^{M}\ket{n}\bra{n}$ denotes the projector onto the $M$-photon subspace, being $\ket{n}$ a Fock state. In doing so, now the eigendecomposition of $\rho^\mu_{[g(\theta)], M}$ can be easily obtained numerically. For later convenience, we will denote the eigendecomposition of the numerator of the RHS of Eq.~(\ref{eq:discrete_proy}) as 
\begin{equation}
\label{eq:eigen_discrete}
\Pi_M\rho^\mu_{[g(\theta)]}\Pi_M=\sum_{n=0}^{M} q_{n| \mu, g(\theta)}{\hat P}(|\varphi_{n, \mu, g(\theta)}\rangle).
\end{equation}

Importantly, this technique also allows to transform the infinite-dimensional SDPs given by Eqs.~\eqref{eq:SDP_yield_inf}-\eqref{eq:SDP_error_inf} onto finite-dimensional SDPs that can be solved numerically. The resulting SDPs and their derivation are provided in Appendix~\ref{app:bounds_SDP}. 

\subsection{Lower bound on the probabilities $p_{n| s, g(\theta)}$}

As explained in the previous subsection, because the states $\rho^\mu_{[g(\theta)]}$ are infinite-dimensional, it might be difficult to calculate their eigendecomposition, and thus the probabilities $p_{n| s, g(\theta)}$. Instead, here we provide a lower bound on these probabilities based on the eigendecomposition given by Eq.~\eqref{eq:eigen_discrete}. In particular, in  Appendix~\ref{app:bounds_SDP} it is shown that
\begin{equation}\label{wed}
p_{n| s, g(\theta)}\geq q_{n| s, g(\theta)}-\epsilon_s:=p_{n| s, g(\theta)}^\text{L}
\end{equation}
with $\epsilon_s=2 \sqrt{1-\operatorname{Tr}\left[\Pi_M \rho^s_{[g(\theta)]} \Pi_M\right]}$.

\section{Simulation results}\label{numerical}

In this section, we now evaluate the secret key rate obtainable for various examples of functions $g(\theta)$. For illustration purposes, we consider three main scenarios, depending on whether or not the function $g(\theta)$ is fully characterized. Also, for the simulations, we consider a simple channel model whose transmission efficiency is given by $10^{-\frac{\gamma}{10}}$, where $\gamma$ (measured in dB) represents the overall system loss, {\it i.e.}, it also includes the effect of the finite detection efficiency of Bob's detectors. Moreover, for simplicity, we disregard any misalignment effect, and assume that the only source of errors are the dark counts of Bob’s detectors, whose rate is set to $p_{d}=10^{-8}$. In addition, as already mentioned, we consider that the BB84 encoding operators are ideal even though the analysis presented here is applicable if this condition is not met, and we take an error correction efficiency $f=1.16$. 

To obtain the bounds $Y_{n, s, g(\theta)}^{Z, \mathrm{L}}$ and $e_{n, s, g(\theta)}^{\mathrm{U}}$ we use the finite-dimensional versions of the SDPs above, which are presented in Appendix~\ref{app:bounds_SDP}. Note that, the resulting secret key rate is an increasing function of $M$. However, the time required to numerically solve such SDPs grows rapidly with this parameter. For this reason, we have set a sufficiently large $M$ so that an increase in this parameter would result in a negligible improvement of the secret key rate as tested numerically.

\subsection{Fully-characterized $g(\theta)$}\label{known_numerical}

Here, we consider the scenario in which the function $g(\theta)$ is completely characterized, and we evaluate two specific examples of practical interest. The first example corresponds to the scenario given by Eq.~(\ref{eq:dis_perfect}), which has been considered in~\cite{Lo_Ma}, while the second example can be interpreted as a noisy version of the first one.

\subsubsection{Ideal discrete phase randomization}\label{i_disc}

\begin{figure}[h]
\centering\includegraphics [width= 8.6cm, height=6cm] {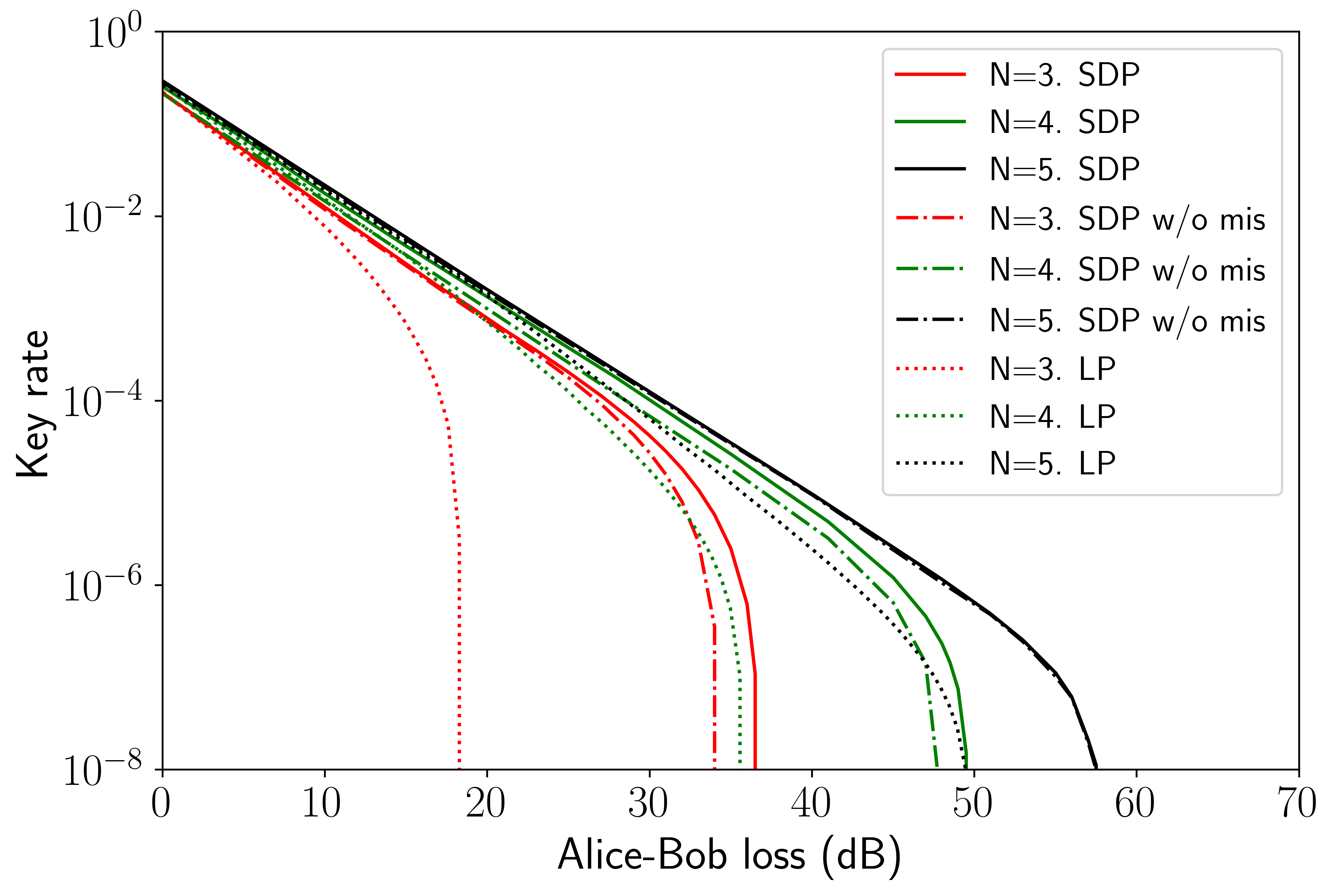}
\caption{\label{fig:comparative} Secret key rate in logarithmic scale versus the overall system loss for the ideal discrete phase-randomization scenario given by Eq.~(\ref{eq:dis_perfect}), as a function of the total number of random phases $N$ selected by Alice. The solid lines correspond to the parameter estimation procedure based on SDP and basis mismatched events considered in this work, while the dashed-dotted lines represent the same procedure overlooking basis mismatched events. Finally, the dotted lines correspond to the analysis in~\cite{Lo_Ma} using linear programming.}
\end{figure}

The results are shown in~Fig.~\ref{fig:comparative} for different values of the total number of random phases $N$ selected by Alice. In particular, the solid lines in the figure have been obtained using the parameter estimation procedure presented in Sec.~\ref{Parameter} based on SDP and the use of basis mismatched events. If we discard these latter events, the obtainable key rate decreases, as illustrated by the dashed-dot lines. Finally, the dotted lines correspond to the analysis in~\cite{Lo_Ma}. For completeness, this latter approach is summarized in Appendix~\ref{app:bounds_LP}. In the first two cases, for simplicity, we set the intensity settings to the possibly sub-optimal values $\omega=0$, $\nu=s/5$ and we optimize $s$ as a function of the overall system loss $\gamma$, while in the later case we set $\omega=0$ and optimize both $\nu$ and $s$ as a function of $\gamma$ (which provides the optimal solution for this approach). Importantly, despite this fact, Fig.~\ref{fig:comparative} shows that the use of SDP and basis mismatched events significantly improve the secret key rate when compared to the results in~\cite{Lo_Ma}. Furthermore, we find that the improvement of using basis mismatched events is more advantageous when $N$ is small. Indeed, when $N\geq5$, this enhancement in performance is almost negligible. This is expected as basis mismatched events do not improve the estimation in the case of ideal continuous phase randomization, $i.e.$, in the limit $N\rightarrow\infty$. On the other hand, when $N$ is small, the eigenstates $|\psi_{n, s, g(\theta)}\rangle$ for $n=0,1$ deviate more from a perfect Fock state, meaning that the virtual states $|\lambda^{\mathrm{virtual}}_{ \Delta, n, s, g(\theta)}\rangle$ deviate more from the $X$-encoded states $\hat{V}_{\Delta_X}|\psi_{n, s, g(\theta)}\rangle$ and thus basis mismatched events provide a tighter estimation.

\begin{figure}[H]
\centering\includegraphics [width= 8.6cm, height=6cm] {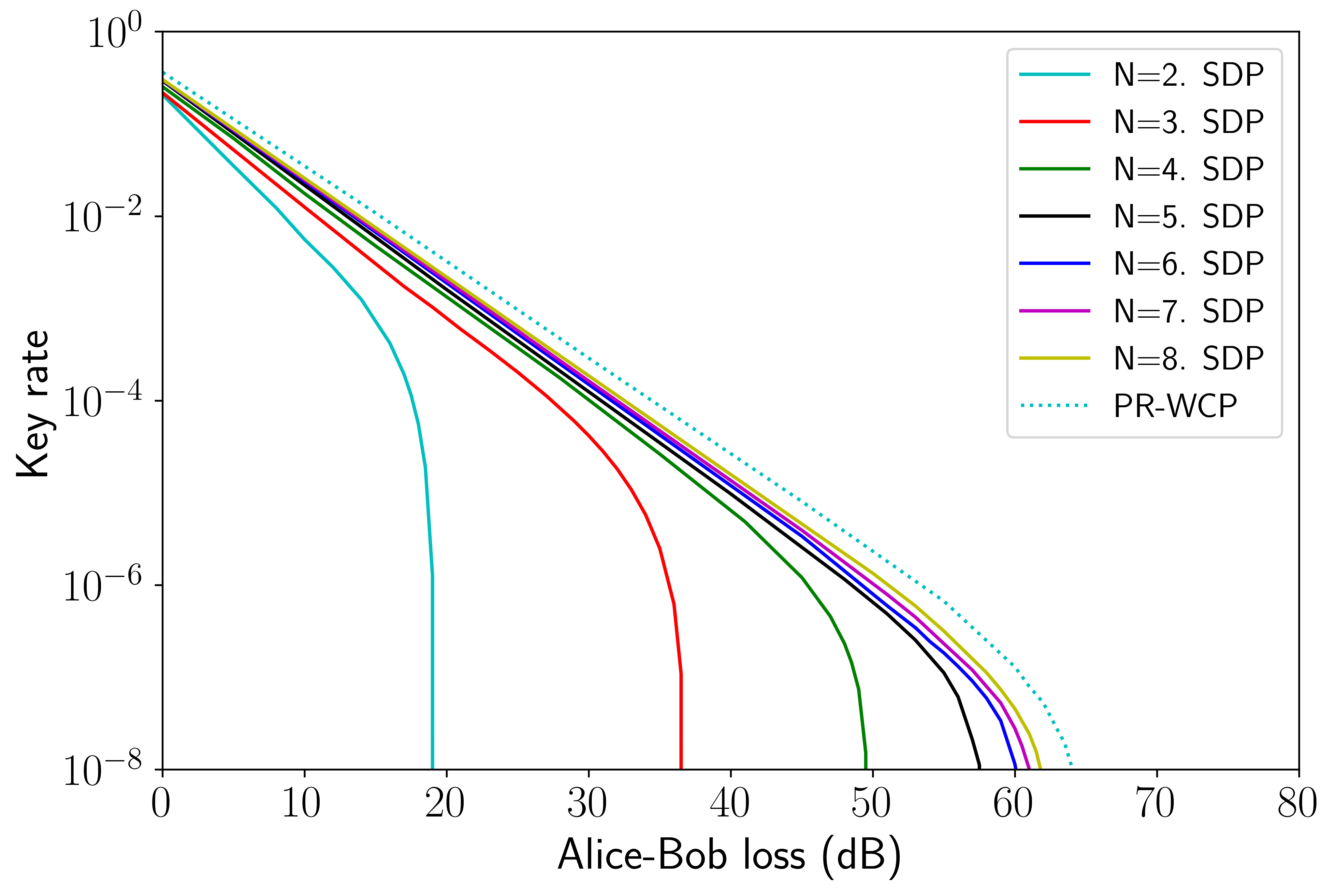}
\caption{\label{fig:fixed} Secret key rate in logarithmic scale versus the overall system loss for the ideal discrete phase-randomization scenario given by Eq.~(\ref{eq:dis_perfect}), as a function of the total number of random phases $N$ selected by Alice, when Alice and Bob employ the parameter estimation procedure based on SDP and basis mismatched events considered in this work. Remarkably, as shown in the figure, only eight random phases are enough to deliver a secret key rate already quite close to the ideal scenario of perfect PR-WCPs, where the phase of each pulse is uniformly random in $[0,2\pi)$.}
\end{figure}

Remarkably, as shown in Fig.~\ref{fig:fixed}, when $N=8$ the secret key rate provided by our approach is already quite close to the ideal scenario where $\theta$ is uniformly random in $[0,2\pi)$. Note that this configuration requires only three random bits per pulse to select the random phase, which does not significantly increase the consumption of the random numbers.
 
\subsubsection{Noisy discrete phase randomization}

Here we consider the situation in which the actual phase encoded by Alice in each emitted pulse follows a certain PDF around the selected discrete value $\theta_{k}=2 \pi k/N$. This might happen due to device imperfections of the phase modulator or the electronics that control it. For concreteness and illustration purposes, we shall assume that this PDF is a truncated Gaussian distribution, though we remark that our analysis can be applied to any given distribution. A truncated Gaussian distribution has the form
\begin{equation}\label{pdf_trun}
f\left(\theta; \theta_k, \sigma_k, \lambda_k, \Lambda_k \right)= \frac{\phi\left(\theta; \theta_k, \sigma_k^2 \right)}{\Phi\left(\Lambda_k; \theta_k, \sigma_k^2 \right)-\Phi\left(\lambda_k; \theta_k, \sigma_k^2 \right)}, \quad
\end{equation}
when the phase $\theta$ is in the interval $\lambda_k<\theta<\Lambda_k$, and zero otherwise. The functions $\phi\left(x; \gamma, \sigma^2 \right)$ and $\Phi\left(x; \gamma, \sigma^2 \right)$ in Eq.~(\ref{pdf_trun}) are, respectively, given by
\begin{equation}
\begin{aligned}
&\phi\left(x; y, z \right)=\frac{1}{\sqrt{2 \pi z}} e^{-\frac{(x-y)^2}{2 z}}, \\
&\Phi\left(x; y, z \right)=\int_{-\infty}^x \frac{1}{\sqrt{2 \pi z}} e^{-\frac{(t-y)^2}{2 z}} d t .
\end{aligned}
\end{equation}
That is, in this scenario the function $g(\theta)$ has the following form
\begin{equation}
\label{func_gauss}
g(\theta)= \frac{1}{N}\sum_{k=0}^{N-1} f\left(\theta; \theta_{k}, \sigma_k, \lambda_{k}, \Lambda_{k} \right)
\end{equation}
for certain parameters $\theta_{k}$, $\sigma_k$, $\lambda_{k}$ and $\Lambda_{k}$.

In the limit when the standard deviations $\sigma_k \to 0$ $\forall k$, Eq.~(\ref{func_gauss}) converges to the PDF given by Eq.~\eqref{eq:dis_perfect}, because in that regime each truncated Gaussian distribution approaches the Dirac delta function. On the other hand, when $\sigma_k \to\infty$, and given that the concatenation of the truncation intervals defined by $\lambda_k$ and $\Lambda_k$ allow the phase to take any value within the range of $[0,2\pi)$ but do not overlap each other, Eq.~(\ref{func_gauss}) converges to the PDF of a uniform distribution in $[0,2\pi)$. Importantly, this means that the achievable secret key rate will increase with higher values of $\sigma_k$, or, to put it in other words, when the uncertainty about the phase actually imprinted by Alice on each of her prepared signals increases, given that $g(\theta)$ is completely characterized. 

\begin{figure}[H]
\centering\includegraphics [width= 8.6cm, height=6cm] {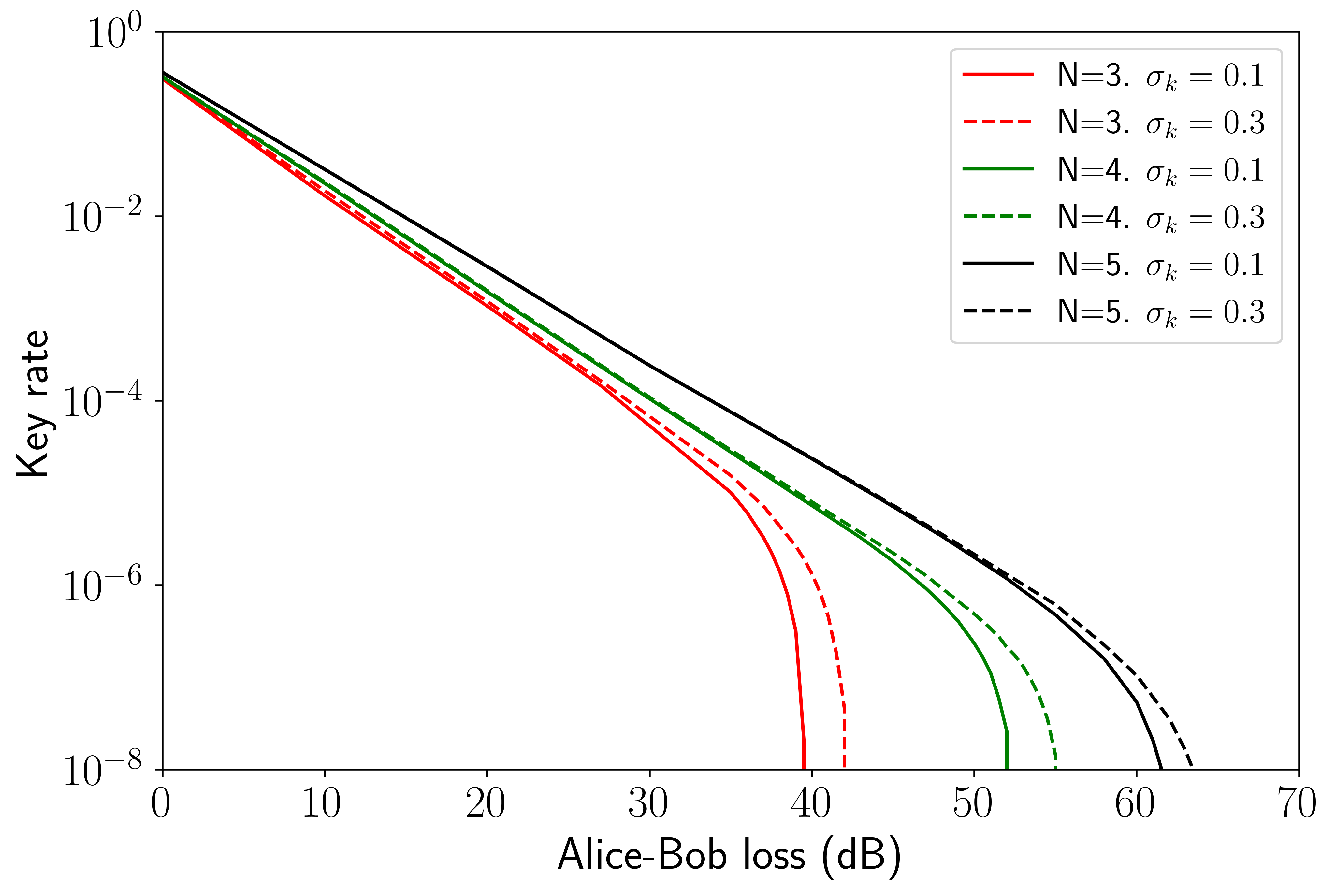}
\caption{\label{fig:gaussian} Secret key rate in logarithmic scale versus the overall system loss when $g(\theta)$ follows the PDF given by Eq.~(\ref{func_gauss}), as a function of the total number of random phases $N$ selected by Alice, and for two different values of the standard deviations $\sigma_k$, which are assumed to be equal for all $k$.
}
\end{figure}

The simulation results are shown in Fig.~\ref{fig:gaussian}, which presents a comparison between the achievable secret key rate for two different values of the standard deviations $\sigma_k$, which, for simplicity, are assumed to be equal for all $k$. As expected, the larger the value of $\sigma_k$ is, the higher the resulting secret key rate, regardless of the number $N$ of random phases selected by Alice, though the improvement is more relevant when $N$ is small. For simplicity and due to the lack of experimental data, Fig.~\ref{fig:gaussian} assumes that $\lambda_{k}=\theta_{k}-3\sigma_k$ and $\Lambda_{k}=\theta_{k}+3\sigma_k$. Moreover, like in the previous example, we set $\omega=0$, $\nu=s/5$ and we optimize $s$ as a function of the overall system loss. 

\subsection{Partially-characterized $g(\theta)$}\label{unknown_numerical}

Here, we now consider the scenario in which only partial information about the function $g(\theta)$ is known. In particular, and for illustration purposes, we shall assume that the actual phase encoded by Alice in each emitted pulse could be any phase within a certain interval around the selected discrete value $\theta_{k}=2 \pi k/N$, but its precise PDF $g(\theta)$ is unknown. Precisely, let $\delta_{\text{max}}$ denote the maximum possible deviation between the actual selected phase $\theta_{k}$ and the actual imprinted phase, which we shall denote by $\hat{\theta}_{k}$. That is, we assume that the actual imprinted phase lies in the interval $\hat{\theta}_{k}\in[\theta_{k}-\delta_{\text{max}},\theta_{k}+\delta_{\text{max}}]$, and we conservatively take the combination of values $\hat{\theta}_{k}$ for all $k$ that minimizes the secret key rate following the analysis presented in Appendix~\ref{app:partially}.

\begin{figure}[h]
\centering\includegraphics [width= 8.6cm, height=6cm] {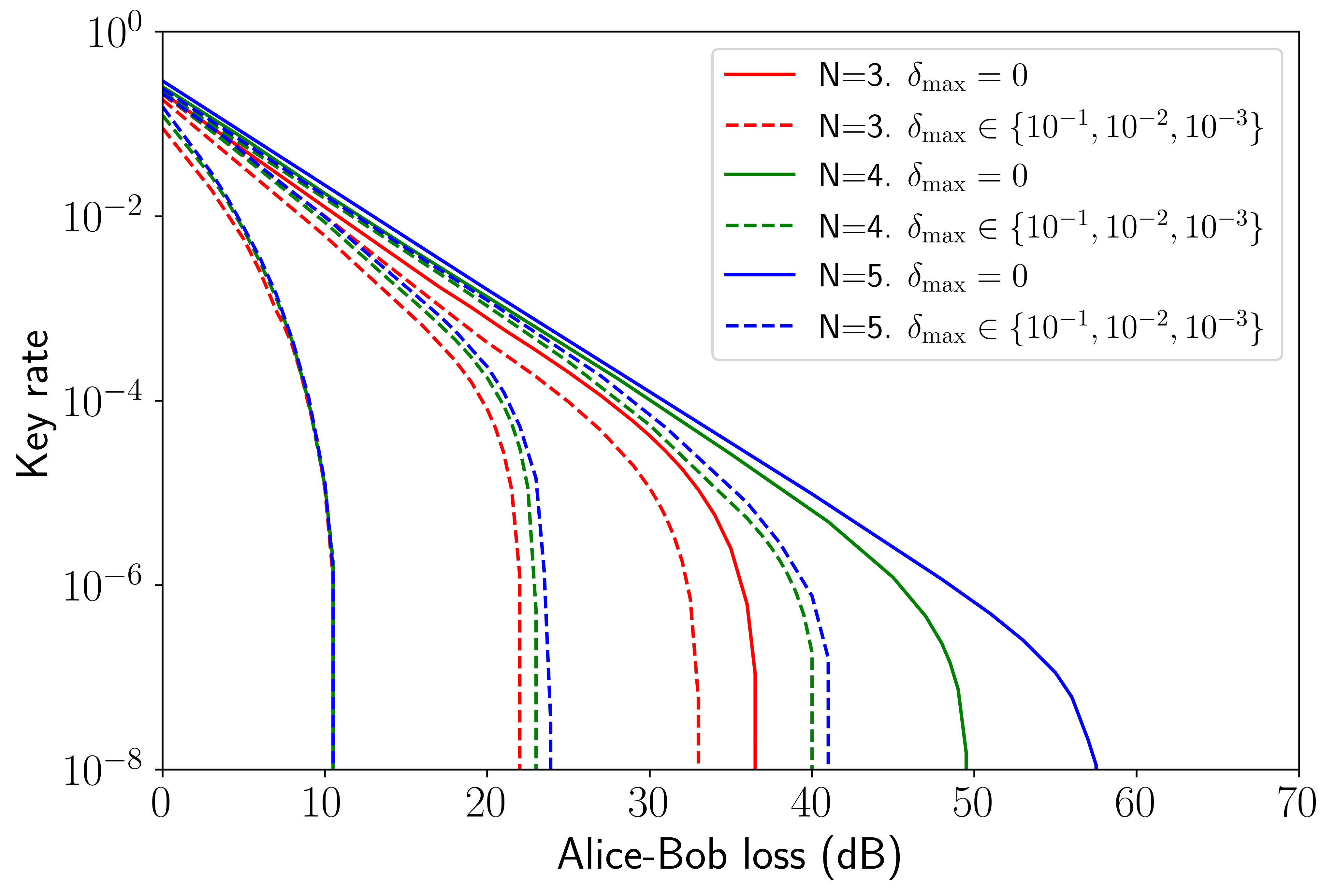}
\caption{\label{fig:unknown} Secret key rate in logarithmic scale versus the overall system loss when the phases lie in the intervals $\theta_{k}\pm\delta_{\text{max}}$ and the function $g(\theta)$ is unknown, as a function of the total number of random phases $N$ selected by Alice and the value of $\delta_{\text{max}}$.}
\end{figure}

The results are illustrated in Fig.~\ref{fig:unknown}, as a function of the total number of phases $N$ selected by Alice and the value of the maximum deviation $\delta_{\text{max}}$. Like in the previous examples, for simplicity,  we fix $\omega=0$, $\nu=s/5$ and we optimize $s$ as a function of the overall system loss. As expected, the larger the value of $\delta_{\text{max}}$ is, the lower the resulting secret key rate. 

Also, from Fig.~\ref{fig:unknown} we see that for higher values of $\delta_{\text{max}}$, the secret key rate becomes less sensitive to the parameter $N$. Indeed, when $\delta_{\text{max}}=10^{-1}$, the achievable secret key rate for the cases $N=3, 4, 5$ essentially overlap each other, which is the left-most curve. This seems to be due to the fact that a significant increase in $\delta_{\text{max}}$ allows in principle for some phases to lie close to each other, or even become identical if this parameter is large enough. Under this situation, the increase of $N$ does not help to improve the performance, as the effective randomness remains almost the same.

\section{Conclusion}\label{conc}

In this paper we have considered the security of decoy-state quantum key distribution (QKD) when the phase of each generated signal is not uniformly random, as requested by the theory, but follows an arbitrary, continuous or discrete, probability density function (PDF). This might happen due to the presence of device imperfections in the phase-randomization process, and/or due to the use of an external phase modulator to imprint the random phases on the generated pulses, which limits the possible selected phases to a finite set. 

Our analysis combines a novel parameter estimation technique, based on semi-definite programming, with the use of basis mismatched events, to tightly estimate the relevant parameters that are needed to evaluate the achievable secret key rate. In doing so, we have shown that decoy-state QKD is rather robust to faulty phase-randomization, particularly when the PDF that governs the random phases is well-characterized. Moreover, our results significantly outperform those of previous works while being also more general, in the sense that they can handle more realistic and practical scenarios. 

This work might be relevant as well to other quantum communication protocols beyond QKD that use laser sources and decoy states.

\section{Acknowledgements}\label{Acknowledgements}

This work was supported by Cisco Systems Inc., the Galician Regional Government (consolidation of Research Units: AtlantTIC), the Spanish Ministry of Economy and Competitiveness (MINECO), the Fondo Europeo de Desarrollo Regional (FEDER) through the grant No. PID2020-118178RB-C21, MICIN with funding from the European Union NextGenerationEU (PRTR-C17.I1) and the Galician Regional Government with own funding through the “Planes Complementarios de I+D+I con las Comunidades Autónomas” in Quantum Communication, the European Union’s Horizon Europe Framework Programme under the Marie Skłodowska-Curie Grant No. 101072637 (Project QSI) and the project "Quantum Security Networks Partnership" (QSNP, grant agreement No 101114043). X.S. acknowledges support from an FPI predoctoral scholarship granted by the Spanish Ministry of Universities. G.C.-L. acknowledges support from JSPS Postdoctoral Fellowships for Research in Japan. K.T. acknowledges support from JSPS KAKENHI Grant Number JP18H05237.

\appendix

\section{Derivation of the SDPs given by Eqs.~\eqref{eq:SDP_yield_inf}-\eqref{eq:SDP_error_inf}}\label{app:infinite}

In this Appendix, we follow the approach in~\cite{Guillermo}, to derive the SDPs presented in Eqs.~\eqref{eq:SDP_yield_inf}-\eqref{eq:SDP_error_inf} of the main text, under the assumption of collective attacks. 

Let $\Omega$ denote a quantum channel (or the action of Eve) that acts independently on each optical pulse emitted by Alice. Also, let us assume that in a certain round, Bob measures the incoming signal with a positive operator valued measure (POVM) that contains the element $\Pi$. In this scenario, the probability that Bob obtains the outcome associated with the element $\Pi$ given that Alice sends him a quantum state $\sigma$ can be expressed as
\begin{eqnarray}
\label{eq:firstA}
\operatorname{Tr}[\Omega(\sigma) \Pi]&=&\operatorname{Tr}\left(\sum_k A_k \sigma A_k^{\dagger} \Pi\right)=\operatorname{Tr}\left(\sigma \sum_k A_k^{\dagger} \Pi A_k\right)\nonumber\\
&=&\operatorname{Tr}(\sigma H),
\end{eqnarray}
where $\Omega(\sigma)$ represents the action of $\Omega$ on $\sigma$, $\left\{A_k\right\}$ denotes the set of Kraus operators corresponding to the operator-sum representation of the channel $\Omega$, and
\begin{equation}
0 \leq H=\sum_k A_k^{\dagger} \Pi A_k \leq \sum_k A_k^{\dagger} A_k=\mathbb{I}.
\end{equation}

Bob measures the incoming signals in either the $Z$ or the $X$ basis. Let us denote the POVM elements corresponding to each of these two measurements by $\left\{\Pi_{0_Z}, \Pi_{1_Z}, \Pi_f\right\}$ and $\left\{\Pi_{0_X}, \Pi_{1_X}, \Pi_f\right\}$, respectively. That is, $\Pi_{b_\alpha}$ represents the POVM element associated to the outcome $b$ in the basis $\alpha$, with $\alpha\in\{Z,X\}$, and $\Pi_{f}$ represents the POVM element associated to an inconclusive outcome. Note that here we are implicitly considering the basis-independent detection efficiency assumption, which means that the POVM element $\Pi_{f}$ is equal for both basis. Let $\Pi_{d}=\mathbb{I}-\Pi_{f}=\Pi_{0_Z}+\Pi_{1_Z}=\Pi_{0_X}+\Pi_{1_X}$ denote the operator associated to a conclusive outcome at Bob's side. Then, after substituting in Eq.~(\ref{eq:firstA}) the state $\sigma$ with Alice's emitted state when she chooses the $Z$ basis, 
\begin{equation}
\rho^{\mu, Z}_{[g(\theta)]}=\frac{1}{2} \hat{V}_{0_Z} \rho^\mu_{[g(\theta)]} \hat{V}_{0_Z}^{\dagger}+\frac{1}{2} \hat{V}_{1_Z} \rho^\mu_{[g(\theta)]} \hat{V}_{1_Z}^{\dagger}, 
\end{equation}
and the operator $\Pi$ with $\Pi_d$, we obtain
\begin{eqnarray}
Q_{\mu, g(\theta)}^Z&=&\operatorname{Tr}[\Omega(\rho^{\mu, Z}_{[g(\theta)]}) \Pi_d]=\operatorname{Tr}[\rho^{\mu, Z}_{[g(\theta)]} H] \nonumber \\
&=&\operatorname{Tr}[\rho^{\mu}_{[g(\theta)]} J_{Z}],\quad
\end{eqnarray}
with $H=\sum_k A_k^{\dagger} \Pi_d A_k$, and the operator $J_{Z}$ satisfying
\begin{equation}
0 \leq J_{Z}=\frac{1}{2}\left(\hat{V}_{0_Z}^{\dagger} H \hat{V}_{0_Z}+\hat{V}_{1_Z}^{\dagger} H \hat{V}_{1_Z}\right) \leq \mathbb{I} .
\end{equation}

Finally, by taking into account that the yield associated to the states $|\psi_{n, s, g(\theta)}\rangle$ encoded in the Z basis is given by 
\begin{eqnarray}
Y_{n, s, g(\theta)}^{Z}&=&\operatorname{Tr}\{\Omega[{\hat P}(|\psi^Z_{n, s, g(\theta)}\rangle)]\Pi_d\} \nonumber \\
&=&\operatorname{Tr}[{\hat P}(|\psi_{n, s, g(\theta)}\rangle) J_{Z}],
\end{eqnarray}
with 
\begin{eqnarray}
{\hat P}(|\psi^Z_{n, s, g(\theta)}\rangle)&=&\frac{1}{2} \hat{V}_{0_Z} {\hat P}(|\psi_{n, s, g(\theta)}\rangle) \hat{V}_{0_Z}^{\dagger} \nonumber \\
&+&\frac{1}{2} \hat{V}_{1_Z} {\hat P}(|\psi_{n, s, g(\theta)}\rangle) \hat{V}_{1_Z}^{\dagger}, 
\end{eqnarray}
we obtain the SDP presented in Eq.~\eqref{eq:SDP_yield_inf}.

Regarding the SDP given by Eq.~\eqref{eq:SDP_error_inf} to estimate the phase error rate, we note that the numerator of Eq.~\eqref{eq:phase_virtual}, can be expressed as
\begin{eqnarray}
&&p^{\mathrm{virtual}}_{\Delta, n, s, g(\theta)}Y_{\Delta, n, s, g(\theta)}^{ (\Delta \oplus 1)_X, \mathrm{virtual}}\nonumber \\
&&=p^{\mathrm{virtual}}_{\Delta, n, s, g(\theta)} \operatorname{Tr}\left\{\Omega[{\hat P}(|\lambda^{\mathrm{virtual}}_{ \Delta, n, s, g(\theta)}\rangle)] \Pi_{(\Delta \oplus 1)_X}\right\} \nonumber \\
&&=\operatorname{Tr}\left[{\hat P}(|\bar{\lambda}^{\mathrm{virtual}}_{\Delta, n, s, g(\theta)}\rangle) L_{(\Delta \oplus 1)_X}\right],
\end{eqnarray}
where $0 \leq L_{(\Delta \oplus 1)_X}=\sum_k A_k^{\dagger} \Pi_{(\Delta \oplus 1)_X} A_k \leq \mathbb{I}$ according to Eq.~(\ref{eq:firstA}), and $|\bar{\lambda}^{\mathrm{virtual}}_{\Delta, n, s, g(\theta)}\rangle=\sqrt{p^{\mathrm{virtual}}_{\Delta, n, s, g(\theta)}}|\lambda^{\mathrm{virtual}}_{ \Delta, n, s, g(\theta)}\rangle$.

By using again Eq.~(\ref{eq:firstA}), we have that the gains $Q_{\mu, g(\theta), b_{\alpha}}^{(\Delta \oplus 1)_X}$ can be expressed as
\begin{equation}
Q_{\mu, g(\theta), b_{\alpha}}^{(\Delta \oplus 1)_X}=\operatorname{Tr}\left[\hat{V}_{b_{\alpha}} \rho_{[g(\theta)]}^{\mu} \hat{V}_{b_{\alpha}}^{\dagger} L_{(\Delta\oplus 1)_X}\right].
\end{equation}

Putting it all together, we find that the SDP presented in Eq.~\eqref{eq:SDP_error_inf} of the main text, provides an upper bound on $p^{\mathrm{virtual}}_{\Delta, n, s, g(\theta)}Y_{\Delta, n, s, g(\theta)}^{(\Delta \oplus 1)_{X},\mathrm{virtual}}$.

\section{Finite-dimensional SDPs when $g(\theta)$ is fully characterized}\label{app:bounds_SDP}

\subsection{Lower bound on the yields $Y_{n, s, g(\theta)}^{Z}$}\label{append_Y}

In this Appendix, we show how to obtain a finite-dimensional relaxation of the SDP given by Eq.~(\ref{eq:SDP_yield_inf}) to find a lower bound on the yields $Y_{n, s, g(\theta)}^{Z}$. For this, we follow again the approach presented in~\cite{Nahar, Guillermo}. The key idea is rather simple: instead of considering the infinite-dimensional state $\rho^{\mu}_{[g(\theta)]}$ given by Eq.~(\ref{eq:diagonal}), we employ a projection $\rho_{[g(\theta)], M}^{\mu}$ of this state onto a finite-dimensional subspace with up to $M$ photons (see Eq.~(\ref{eq:discrete_proy})), and then we relax the original constraints of the SDP accordingly.

We begin by briefly introducing some helpful results for this purpose. The first one is a direct consequence of the Cauchy-Schwarz inequality in Hilbert spaces~\cite{Lo_Preskill,Pereira2020}, which allows to relate the quantities $\operatorname{Tr}[\sigma H]$ and $\operatorname{Tr}[\rho H]$, with $0 \leq H \leq \mathbb{I}$, as a function of the fidelity between the states $\sigma$ and $\rho$, 
\begin{equation}
F(\rho,\sigma)=\operatorname{Tr}\left[\sqrt{\sqrt{\sigma} \rho \sqrt{\sigma}}\right]^2. 
\end{equation}
In particular, it states that
\begin{equation}
\label{eq:CS}
\begin{aligned}
&G_{-}\left(\operatorname{Tr}\left[\rho H\right], F\left(\sigma, \rho\right)\right) \leq \operatorname{Tr}[\sigma H]\\
&\leq G_{+}\left(\operatorname{Tr}\left[\rho H\right], F\left(\sigma, \rho\right)\right),
\end{aligned}
\end{equation}
with the functions $G_{\pm}(y, z)$ being defined as
\begin{equation}
G_{-}(y, z)= \begin{cases}g_{-}(y, z) & \text { if } y>1-z \\ 0 & \text { otherwise },\end{cases}
\end{equation}
and 
\begin{equation}
G_{+}(y, z)= \begin{cases}g_{+}(y, z) & \text { if } y<z \\ 1 & \text { otherwise },\end{cases}
\end{equation}
with $g_{\pm}(y, z)=y+(1-z)(1-2 y) \pm 2 \sqrt{z(1-z) y(1-y)}$.

The remaining results we use, {\it i.e.} Eqs.(\ref{one})-(\ref{second})-(\ref{eq:Fvec})-(\ref{third}) below, have been derived in~\cite{Nahar, Guillermo, Winter}. In particular, we have that  
\begin{equation}\label{one}
\begin{aligned}
&F\left(\rho_{[g(\theta)]}^{\mu}, \rho_{[g(\theta)], M}^{\mu}\right)=\operatorname{Tr}\left[\Pi_M \rho_{[g(\theta)]}^\mu \Pi_M\right]\\
&=\sum_{n=0}^M q_{n|s,g(\theta)}:=F^{\text{proj}}_{\mu,g(\theta)},
\end{aligned}
\end{equation}
where the coefficients $q_{n|s,g(\theta)}$ are given in Eq.~(\ref{eq:eigen_discrete}).  Also, we have that the quantities $\left|p_{n \mid \mu, g(\theta)}-q_{n \mid \mu,g(\theta)}\right|$ can be upper bounded as
\begin{eqnarray}\label{second}
\left|p_{n \mid \mu, g(\theta)}-q_{n \mid \mu,g(\theta)}\right| &\leq& 2 \sqrt{1-\operatorname{Tr}\left[\Pi_M \rho_{[g(\theta)]}^\mu \Pi_M\right]}\nonumber \\
&=&2 \sqrt{1-F^{\text{proj}}_{\mu,g(\theta)}}=: \epsilon_{\mu}.
\end{eqnarray}
Finally, the fidelity $F\left({\hat P}(|\varphi_{n,\mu,g(\theta)}\rangle),{\hat P}(|\psi_{n,\mu,g(\theta)}\rangle) \right)=|\langle\varphi_{n,\mu,g(\theta)}|\psi_{n,\mu,g(\theta)}\rangle|^2$
satisfies
\begin{eqnarray}
\label{eq:Fvec}
F\left({\hat P}(|\varphi_{n,\mu,g(\theta)}\rangle),{\hat P}(|\psi_{n,\mu,g(\theta)}\rangle) \right) 
&\geq& 1-\left(\frac{\epsilon_{\mu}}{\delta_{n,\mu}}\right)^2 \nonumber \\
&:=&F_{n,\mu,g(\theta)}^{\text{vec}},
\end{eqnarray}
with 
\begin{eqnarray}\label{third}
\delta_{0,\mu}&=&q_{0\mid \mu, g(\theta)}-q_{1\mid \mu, g(\theta)}-\epsilon_{\mu} \nonumber \\
\delta_{n,\mu}&=&\min \big\{q_{n-1 \mid \mu, g(\theta)}-q_{n \mid \mu, g(\theta)}-\epsilon_{\mu}, q_{n \mid \mu, g(\theta)}\nonumber \\
&-&q_{n+1 \mid \mu, g(\theta)}-\epsilon_{\mu}\big\}.
\end{eqnarray}

Then,  from Eqs.~(\ref{sat_morn1})-\eqref{eq:CS}-(\ref{eq:Fvec}) we have that
\begin{eqnarray}
\label{eq:first}
Y_{n, s, g(\theta)}^{Z, {\rm L}}&=&\operatorname{Tr}\left[{\hat P}(|\psi_{n, s, g(\theta)}\rangle)J_{Z}^*\right]\nonumber\\
&\geq&G_{-}\left(\operatorname{Tr}\left[{\hat P}(|\varphi_{n, s, g(\theta)}\rangle)J_{Z}^*\right], F_{n,s,g(\theta)}^{\text {vec}}\right),\quad
\end{eqnarray}
where $J_{Z}^{*}$ is the solution to the SDP presented in Eq.~\eqref{eq:SDP_yield_inf}, and we have used the fact that $G_{-}$ is increasing with respect to its second argument. Since $G_{-}(y,z)$ is decreasing with respect to its first argument, one can lower bound Eq.~(\ref{eq:first}) by finding a lower bound on its first argument. 

From Eq.~(\ref{eq:CS}), we have that
 
\begin{eqnarray}
G_{-}\left(Q_{\mu, g(\theta)}^Z, F^{\text{proj}}_{\mu,g(\theta)}\right) &\leq& \operatorname{Tr}\left[\rho_{[g(\theta)],M}^{\mu} J_{Z}\right]\nonumber \\
&\leq& G_{+}\left(Q_{\mu, g(\theta)}^Z, F^{\text{proj}}_{\mu,g(\theta)}\right),\quad\quad\quad 
\end{eqnarray}
with the operator $J_{Z}$ defined in Eq.~(\ref{eq:SDP_yield_inf}). Here, since the states $\rho_{[g(\theta)],M}^{\mu}$ are finite dimensional, the calculation of $ \operatorname{Tr}\left[\rho_{[g(\theta)],M}^{\mu} J_{Z}\right]$ can be restricted to operators $J_{Z}$ that act on their finite subspace. Putting it all together, we find that a lower bound on $Y_{n, s, g(\theta)}^{Z}$ can be obtained by solving the following finite-dimensional SDP program
\begin{equation}
\label{eq:sdpfinite}
\begin{array}{ll}
\min _{J_{Z}} & \operatorname{Tr}\left[ {\hat P}(|\varphi_{n, s, g(\theta)}\rangle) J_{Z}\right] \\
\text { s.t. } & G_{-}\left(Q_{\mu, g(\theta)}^Z, F^{\text{proj}}_{\mu,g(\theta)}\right) \leq \operatorname{Tr}\left[\rho_{[g(\theta)],M}^{\mu} J_{Z}\right]\\
&\leq G_{+}\left(Q_{\mu, g(\theta)}^Z, F^{\text{proj}}_{\mu,g(\theta)}\right), \quad \forall \mu\in\{s, \nu, \omega\} \\
& 0 \leq J_{Z} \leq \mathbb{I}.
\end{array}
\end{equation}
That is, we have that 
\begin{equation}
\label{eq:trivial}
\operatorname{Tr}\left[{\hat P}(|\varphi_{n, s, g(\theta)}\rangle)J_{Z}^*\right]\geq \operatorname{Tr}\left[{\hat P}(|\varphi_{n, s, g(\theta)}\rangle)J_{Z}^{**}\right],
\end{equation}
with $J_{Z}^{**}$ being the solution to the SDP in Eq.~\eqref{eq:sdpfinite}, and $J_{Z}^*$ the solution to Eq.~\eqref{eq:SDP_yield_inf}. This holds because the constrains in Eq.~\eqref{eq:sdpfinite} are looser than those in Eq.~\eqref{eq:SDP_yield_inf}.

Finally, by combining Eq.~(\ref{eq:first}) with Eq.~\eqref{eq:trivial} we have that
\begin{eqnarray}
\label{eq:firstB}
Y_{n, s, g(\theta)}^{Z, {\rm L}}&\geq&G_{-}\left(\operatorname{Tr}\left[{\hat P}(|\varphi_{n, s, g(\theta)}\rangle)J_{Z}^{**}\right], F_{n,s,g(\theta)}^{\text {vec}}\right) \nonumber \\
&:=&{\tilde Y}_{n, s, g(\theta)}^{Z, {\rm L}}.
\end{eqnarray}
The lower bound ${\tilde Y}_{n, s, g(\theta)}^{Z, {\rm L}}$ is the one we use in our simulations in Sec.~\ref{known_numerical}.

\subsection{Upper bound on the phase-error rates $e_{n, s, g(\theta)}$} \label{sec:finite_error}

In this Appendix, we show how to estimate an upper bound on $e_{n, s, g(\theta)}$ by using a finite-dimensional SDP. To do so, let us also define the operator
\begin{equation}
M_{\mathrm{ph}}:=|0_X\rangle\langle 0_X|\otimes L_{1_X}^{*}+| 1_X\rangle\langle 1_X| \otimes L_{0_X}^{*},
\end{equation}
where $L_{(\Delta \oplus 1)_X}^{*}$ denotes the solution to the SDP given by Eq.~(\ref{eq:SDP_error_inf}), so that 
\begin{eqnarray}
&&\sum_{\Delta=0}^1p^{\mathrm{virtual}}_{\Delta, n, s, g(\theta)}Y_{\Delta, n, s, g(\theta)}^{ (\Delta \oplus 1)_X, \mathrm{virtual}}\leq\sum_{\Delta=0}^1\operatorname{Tr}\Big[{\hat P}(|\bar{\lambda}^{\text {virtual}}_{\Delta, n, s, g(\theta)}\rangle)\nonumber \\
&&\times{}L_{(\Delta \oplus 1)_X}^{*}\Big]=\operatorname{Tr}\left[{\hat P}(|\Psi^{Z}_{n, s, g(\theta)}\rangle) M_{\text {ph}}\right].
\end{eqnarray}

Now, let us define the finite-dimensional state
\begin{equation}\label{purif_norm}
|\Psi^{Z, M}_{n, s, g(\theta)}\rangle=\frac{1}{\sqrt{2}}\left(|0_Z\rangle_A \hat{V}_{0_Z}+|1_Z\rangle_A \hat{V}_{1_Z}\right)|\varphi_{n, s, g(\theta)}\rangle, 
\end{equation}
and the unnormalized states $|\bar{\lambda}^{\mathrm{virtual},M}_{ \Delta, n, s, g(\theta)}\rangle$ as 
\begin{eqnarray}
|\bar{\lambda}^{\mathrm{virtual},M}_{ \Delta, n, s, g(\theta)}\rangle&=&{ }_A\langle\Delta_X |\Psi^{Z, M}_{n, s, g(\theta)}\rangle \nonumber\\
&=&\frac{1}{2}\left[\hat{V}_{0_Z}+(-1)^\Delta \hat{V}_{1_Z}\right]|\varphi_{n, s, g(\theta)}\rangle.\quad\
\end{eqnarray}

Then, we have that
\begin{eqnarray}
\left|\langle\Psi^{Z, M}_{n, s, g(\theta)}|\Psi^{Z}_{n, s, g(\theta)}\rangle\right|^2&=&\left|\langle\varphi_{n,s,g(\theta)}|\psi_{n,s,g(\theta)}\rangle\right|^2 \nonumber \\
&\geq&F^{\mathrm{vec}}_{n,s,g(\theta)},
\end{eqnarray}
where we have used Eq.~\eqref{eq:Fvec} and the fact that $\hat{V}_{0 Z}^{\dagger} \hat{V}_{0 Z}=\hat{V}_{1 Z}^{\dagger} \hat{V}_{1 Z} = \mathbb{I}$. Now, by applying the Cauchy-Schwarz constraint given by Eq.~(\ref{eq:CS}), and taking into account the fact that $G_{+}(y,z)$ is a decreasing function with respect to its second argument, we find that
\begin{equation}
\begin{aligned}
&\operatorname{Tr}\left[{\hat P}(|\Psi^{Z}_{n, s, g(\theta)}\rangle) M_{\text {ph}}\right] \leq\\
&G_{+}\left(\operatorname{Tr}\left[{\hat P}(|\Psi^{Z, M}_{n, s, g(\theta)}\rangle) M_{\text {ph}}\right], F^{\text{vec}}_{n,s,g(\theta)}\right).
\end{aligned}
\end{equation}

Importantly, since $G_{+}(y,z)$ is an increasing function with respect to its first argument, one can upper bound the previous equation by finding an upper bound on its first argument. Moreover, since the states $|\Psi^{Z, M}_{n, s, g(\theta)}\rangle$ are finite dimensional, one can restrict the optimization search to operators $L$ that act on the corresponding finite subspace. In particular, we have that
\begin{equation}
\begin{aligned}
&\operatorname{Tr}\left[{\hat P}(|\Psi^{Z, M}_{n, s, g(\theta)}\rangle) M_{\text {ph}}\right]=
\sum_{\Delta=0}^1\operatorname{Tr}\Big[{\hat P}(|\bar{\lambda}^{\text {virtual}, M}_{\Delta, n, s, g(\theta)}\rangle)\\
&\times{}L_{(\Delta \oplus 1)_X}^{*}\Big]\leq{} \sum_{\Delta=0}^1\operatorname{Tr}\left[{\hat P}(|\bar{\lambda}^{\text {virtual}, M}_{\Delta, n, s, g(\theta)}\rangle) L_{(\Delta \oplus 1)_X}^{**}\right].
\end{aligned}
\end{equation}
where $ L_{(\Delta \oplus 1)_X}^{**}$ is the solution to the finite-dimensional SDP presented below.

Likewise, the constraints in Eq.~(\ref{eq:SDP_error_inf}) can be relaxed by using essentially the same techniques discussed in Appendix~\ref{append_Y}. In doing so, we find that an upper bound on $\operatorname{Tr}\left[{\hat P}(|\bar{ \lambda}^{\text {virtual}, M}_{\Delta, n, s, g(\theta)}\rangle) L_{(\Delta \oplus 1)_X}\right]$ can be found by solving the following SDP
\begin{widetext}
\begin{equation}
\label{eq:errorfinite}
\begin{array}{ll}
\max _{L_{(\Delta \oplus 1)_X}} & \operatorname{Tr}\left[{\hat P}(|\bar{ \lambda}^{\text {virtual}, M}_{\Delta, n, s, g(\theta)}\rangle) L_{(\Delta \oplus 1)_X}\right] \\
\text { s.t. } & G_{-}\left(Q_{\mu,g(\theta), b_{\alpha}}^{(\Delta \oplus 1)_X}, F^{\text{proj}}_{\mu,g(\theta)}\right) \leq \operatorname{Tr}\left[\hat{V}_{b_{\alpha}} \rho_{[g(\theta)],\text{M}}^{\mu} \hat{V}_{b_{\alpha}}^{\dagger} L_{(\Delta \oplus 1)_X}\right] \leq G_{+}\left(Q_{\mu,g(\theta), b_{\alpha}}^{(\Delta \oplus 1)_X}, F^{\text{proj}}_{\mu,g(\theta)}\right), \\
&\forall \mu \in \{s, \nu, \omega \}, \forall b\in\{0,1\}, \forall \alpha\in\{Z,X\} \\
& 0 \leq L_{(\Delta \oplus 1)_X} \leq \mathbb{I} ,
\end{array}
\end{equation}
\end{widetext}
where $F^{\text{proj}}_{\mu,g(\theta)}$ is given by Eq.~(\ref{one}). 

Let $L_{(\Delta \oplus 1)_X,}^{**}$ denote the operator that maximizes the SDP given by Eq.~(\ref{eq:errorfinite}), then 
\begin{widetext}
\begin{equation}
e_{n, s, g(\theta)}\leq\frac{1}{{\tilde Y}_{n, s, g(\theta)}^{Z, {\rm L}}}G_{+}\left(\sum_{\Delta=0}^1\operatorname{Tr}\left[{\hat P}(|\bar{\lambda}^{\text {virtual}, M}_{\Delta, n, s, g(\theta)}\rangle) L_{(\Delta \oplus 1)_X}^{**}\right], F^{\text{vec}}_{n,s,g(\theta)}\right)\nonumber \\
:={\tilde e}_{n, s, g(\theta)}^{\mathrm{U}}. 
\end{equation}
\end{widetext}
This is the upper bound that we use in our simulations in Sec.~\ref{known_numerical}.

\section{Finite-dimensional SDPs when $g(\theta)$ is partially characterized}\label{app:partially}

Here, we consider the scenario studied in Sec.~\ref{unknown_numerical}, {\it i.e.}, when the actual imprinted phases lies in certain intervals $\hat{\theta}_{k}\in[\theta_{k}-\delta_{\text{max}},\theta_{k}+\delta_{\text{max}}]$, with $\theta_{k}=2 \pi k/N$, and the exact form of $g(\theta)$ is unknown. 

A direct solution to this case could be found as follows. First, one defines a dense grid with $p$ discrete values within each interval, and then one follows essentially the approach in Sec.~\ref{i_disc} for each possible combination of these discrete phases from the different intervals. The secret key rate would then correspond to the worst case scenario, {\it i.e.}, the one that minimizes it among all possible combinations. The main drawback of this approach is, however, that the number of SDPs that needs to be solved grows very rapidly, as $\propto p^N$. 

Instead, here we introduce a much simpler approach based on a modified version of the SDPs presented in Eqs.~\eqref{eq:sdpfinite}-\eqref{eq:errorfinite}. In particular, let $f(\theta)$ denote the PDF associated to the ideal discrete phase randomization scenario given by Eq.~(\ref{eq:dis_perfect}), and let $\rho^{\mu}_{[f(\theta)], M}$ be the finite-dimensional state obtained by projecting $\rho^{\mu}_{[f(\theta)]}$ onto the subspace that contains up to $M$ photons. Also, let $\rho^{\mu}_{[g(\theta)]}$ denote the state actually emitted by Alice in the scenario described above, {\it i.e.}, when $g(\theta)$ is partially characterized. Then, we can bound the fidelity between $\rho^{\mu}_{[g(\theta)]}$ and $\rho^{\mu}_{[f(\theta)], M}$ by means of the Bures distance, which is defined as~\cite{bures}
\begin{equation}
\label{eq:bures_def}
d_{B}(\rho,\sigma)^{2} = 2[1-\sqrt{F(\rho,\sigma)}],
\end{equation}
for any state $\rho$ and $\sigma$. This distance satisfies the triangle inequality~\cite{bures}, which means that 
\begin{eqnarray}
\label{eq:bures_triangle}
\sqrt{F(\rho^{\mu}_{[g(\theta)]}, \rho^{\mu}_{[f(\theta)], M})}&=&1 - \frac{1}{2} d_{B}(\rho^{\mu}_{[g(\theta)]}, \rho^{\mu}_{[f(\theta)], M})^{2} \nonumber\\
&\geq& 1 - \frac{1}{2}\Big[d_{B} (\rho^{\mu}_{[f(\theta)]},  \rho^{\mu}_{[f(\theta)], M}) \nonumber\\
&+&d_{B}(\rho^{\mu}_{[g(\theta)]}, \rho^{\mu}_{[f(\theta)]})\Big]^{2}.
\end{eqnarray}

We now compute the fidelities that correspond to the Bures distances $d_{B} (\rho^{\mu}_{[f(\theta)]},  \rho^{\mu}_{[f(\theta)], M})$ and $d_{B}(\rho^{\mu}_{[g(\theta)]}, \rho^{\mu}_{[f(\theta)]})$ so that, via Eq.~\eqref{eq:bures_def}, we can obtain the necessary fidelity bound with Eq.~\eqref{eq:bures_triangle}.

In particular, from Eq.~(\ref{one}), we have that $F(\rho^{\mu}_{[f(\theta)]},  \rho^{\mu}_{[f(\theta)], M})=F^{\text{proj}}_{\mu,f(\theta)}$. The fidelity $F(\rho^{\mu}_{g(\theta)}, \rho^{\mu}_{[f(\theta)]})$, on the other hand, can be computed by considering the following purifications of the states $\rho^{\mu}_{[f(\theta)]}$ and $\rho^{\mu}_{[g(\theta)]}$, respectively,
\begin{equation}
\label{eq:purifications}
\begin{aligned}
& |\psi_{[f(\theta)]}^{\mu,N}\rangle=\frac{1}{\sqrt{N}} \sum_{k=0}^{N-1}|k\rangle\left|\sqrt{\mu} e^{2 \pi k i / N}\right\rangle, \\
& |\psi_{[g(\theta)]}^{\mu,N}\rangle=\frac{1}{\sqrt{N}} \sum_{k=0}^{N-1} e^{i \phi_k}|k\rangle\left|\sqrt{\mu} e^{i\left(2 \pi k  / N+\delta_k\right) }\right\rangle.
\end{aligned}
\end{equation}
We find, therefore, that 
\begin{eqnarray}
&&F(\rho^{\mu}_{[g(\theta)]}, \rho_{[f(\theta)]}^{\mu})\geq |\langle\psi_{[f(\theta)]}^{\mu,N}|\psi_{[g(\theta)]}^{\mu,N}\rangle|^{2} \nonumber \\
&&=\Big|\sum_{k=0}^{N-1} \frac{1}{N}\langle\sqrt{\mu}  e^{2 \pi k i / N}| \sqrt{\mu} e^{i(2 \pi k / N+\delta_k)}\rangle\Big|^2 \nonumber \\
&&\geq \Big|\sum_{k=0}^{N-1} \frac{1}{N}\langle\sqrt{\mu}  e^{2 \pi k i / N} |\sqrt{\mu} e^{i(2 \pi k / N+\delta_{\text{max}})}\rangle\Big|^2 \nonumber \\
&&= |\langle \sqrt{\mu}| \sqrt{\mu}e^{i\delta_{\text{max}}}\rangle|^2,
\end{eqnarray}
where in the first inequality we have used the fact that the states on the RHS are a purification of those on the LHS; in the first equality we have taken into account that the phases $\phi_{k}$ in Eq.~\eqref{eq:purifications} can be chosen so that they cancel the phase of the inner product, and in the second inequality we have used the fact that $|\delta_{k}|\leq \delta_{\text{max}}$ $\forall k$.

Since the function $g(\theta)$ is unknown, we do not have access to the exact form of the eigenvectors $\ket{\varphi_{n,s,[g(\theta)]}}$ of $\rho_{[g(\theta)],M}^{s}$ which are needed to solve the relevant finite-dimensional SDP, but we can lower bound the value of $\text{Tr}[{\hat P}(\ket{\varphi_{n,s,[g(\theta)]}})J_{Z}]$, with $0 \leq J_{Z} \leq \mathbb{I}$, by employing the Cauchy-Schwartz constraint presented in Eq.~(\ref{eq:CS}). Precisely, we have that
\begin{eqnarray}
\label{eq:CS_yield}
&&\text{Tr}\Big[{\hat P}(\ket{\varphi_{n,s,[g(\theta)]}})J_{Z}\Big]\geq \\
&&G_{-}\left(\text{Tr}\Big[{\hat P}(\ket{\varphi_{n,s,[f(\theta)]}})J_{Z}\Big], F(\ket{\varphi_{n,s,[g(\theta)]}},\ket{\varphi_{n,s,[f(\theta)]}})\right), \nonumber
\end{eqnarray}
where $\ket{\varphi_{n,s,[f(\theta)]}}$ are the eigenvectors of $\rho_{[f(\theta)],M}^{s}$, and the value of $F(\ket{\varphi_{n,s,[g(\theta)]}},\ket{\varphi_{n,s,[f(\theta)]}})$ is calcuated numerically as explained below.

With these considerations, we can now find a lower bound on the yields $Y_{n, s, g(\theta)}^{Z}$. For this, we first solve the following optimization problem to find the operator $J^{**}_{\text{z}}$ that minimizes its objective function 
\begin{widetext}
\begin{equation}
\label{eq:SDP_unknown}
\begin{array}{cl}
\min_ {J_{Z}} & \text{Tr}\Big[{\hat P}(\ket{\varphi_{n,s,[f(\theta)]}})J_{Z}\Big]\\
\text { s.t. } & G_{-}\left(Q_{\mu, g(\theta)}^Z, F(\rho^{\mu}_{[g(\theta)]},  \rho^{\mu}_{[f(\theta)], M})\right) \leq \text{Tr}\Big[\rho_{[f(\theta)],M}^{\mu}J_{Z}\Big]\leq G_{+}\left(Q_{\mu, g(\theta)}^Z, F(\rho^{\mu}_{[g(\theta)]},  \rho^{\mu}_{[f(\theta)], M})\right),\\
& 0 \leq J_{Z} \leq \mathbb{I}.
\end{array}
\end{equation}
\end{widetext}
Following Eq.~\eqref{eq:CS_yield}, we now define
\begin{equation}
\begin{aligned}
\label{eq:CS_yield_2}
&\hat{Y}_{n, s, g(\theta)}^{Z, {\rm L}}:= \\
&G_{-}\left(\text{Tr}\Big[{\hat P}(\ket{\varphi_{n,s,[f(\theta)]}})J^{**}_{Z}\Big], F(\ket{\varphi_{n,s,[g(\theta)]}},\ket{\varphi_{n,s,[f(\theta)]}})\right).
\end{aligned}
\end{equation}
Finally, by using the arguments introduced in Appendix~\ref{append_Y}, we obtain that a lower bound on $Y_{n, s, g(\theta)}^{Z}$ is given by
\begin{equation}
\label{eq:C8}
Y_{n, s, g(\theta)}^{Z} \geq G_{-}\left(\hat{Y}_{n, s, g(\theta)}^{Z, {\rm L}}, F^{\text{vec}}_{n,s,g(\theta)}\right):=\tilde{Y}_{n, s, g(\theta)}^{Z, {\rm L}}.
\end{equation}

Note that, since we do not know which values of $\hat{\theta}_{k}$ result in the set of states $\ket{\varphi_{n,s,[g(\theta)]}}$ that minimizes the key rate, we find the worst case scenario numerically. To do so, we implement a Montecarlo simulation by considering a dense grid of values in $\theta_{k}\pm \delta_{\text{max}}$ for every $k$ and we find the combination of $\hat{\theta}_{k}$ that minimizes Eq.~\eqref{eq:C8} (which includes the fidelity in Eq.~\eqref{eq:CS_yield_2}). This allow us to find the desired lower bound with arbitrary precision. Also, note that the number of SDPs that need to be solved grows very rapidly in the case of the direct solution mentioned at the beginning of this section. With this approach, this problem has been circumvented by reducing it to a simple calculation of the fidelities, which makes it computationally much faster, despite possibly providing looser bounds.

Regarding the estimation of an upper bound on the phase error rate, we follow the same procedure described in Appendix~\ref{sec:finite_error}. In doing so, we first solve the following finite-dimensional SDP,
\begin{widetext}
\begin{equation}
\label{eq:SDP_unknown_2}
\begin{array}{ll}
\max _{L_{(\Delta \oplus 1)_X}} &  
\operatorname{Tr}\Big[\hat{P}(\ket{\bar{\lambda}_{\Delta, n,s,[f(\theta)]}^{\text{virtual},M}})L_{(\Delta \oplus 1)_X}\Big] \\
\text { s.t. } & G_{-}\left(Q_{\mu, b_{\alpha}}^{(\Delta \oplus 1)_X}, F(\rho^{\mu}_{[g(\theta)]},  \rho^{\mu}_{[f(\theta)], M})\right) \leq \operatorname{Tr}\Big[\hat{V}_{b_{\alpha}} \rho_{[f(\theta)], M}^{\mu} \hat{V}_{b_{\alpha}}^{\dagger} L_{(\Delta \oplus 1)_X}\Big]  \leq G_{+}\left(Q_{\mu, b_{\alpha}}^{(\Delta \oplus 1)_X}, F(\rho^{\mu}_{[g(\theta)]},  \rho^{\mu}_{[f(\theta)], M})\right),  \\
& 0 \leq L_{(\Delta \oplus 1)_X} \leq \mathbb{I},
\end{array}
\end{equation}
\end{widetext}
where $Q_{\mu, b_{\alpha}}^{(\Delta \oplus 1)_X}$ represents the observed rate at which Bob obtains the result $(\Delta \oplus 1)_X$ conditioned on Alice choosing the intensity setting $\mu$, the basis $\alpha$, the bit value $b$ and Bob choosing the $X$ basis. Now, similarly to Eq.~\eqref{eq:CS_yield_2}, we define 
\begin{widetext}
\begin{equation}
\hat{e}^{\rm U}_{n, s, g(\theta)}:=\sum_{\Delta=0}^{1}G_{+}\left(\text{Tr}\Big[\hat{P}(\ket{\bar{\lambda}_{\Delta, n,s,[f(\theta)]}^{\text{virtual},M}})L^{**}_{(\Delta \oplus 1)_X}\Big], F(\ket{\bar{\lambda}_{\Delta, n,s,[f(\theta)]}^{\text{virtual},M}},\ket{\bar{\lambda}_{\Delta,n,s,[g(\theta)]}^{\text{virtual},M}})\right),
\end{equation}
\end{widetext}
where $L^{**}_{(\Delta \oplus 1)_X}$ is the solution to Eq.~\eqref{eq:SDP_unknown_2}. This way, we obtain that the phase error rate $e_{n, s, g(\theta)}$ is upper bounded by 
\begin{equation}
\label{eq:C11}
e_{n, s, g(\theta)}\leq\frac{G_{+}\left(\hat{e}_{n, s, g(\theta)}^{\rm U}, F^{\text{vec}}_{n,s,g(\theta)}\right)}{\tilde{Y}_{n, s, g(\theta)}^{Z, {\rm L}}} :={\tilde e}_{n, s, g(\theta)}^{\mathrm{U}}.
\end{equation}
where again, we use the combination of $\hat{\theta}_{k}$ that maximizes Eq.~\eqref{eq:C11} to obtain the relevant upper bound.

The bounds $\tilde{Y}_{n, s, g(\theta)}^{Z, L}$ and ${\tilde e}_{n, s, g(\theta)}^{\mathrm{U}}$ are used in the simulations presented in Sec.~\ref{unknown_numerical}.

As shown in Fig.~\ref{fig:unknown}, higher values of $\delta_{\text{max}}$ result in an almost negligible impact of the parameter $N$ on the secret key rate, as explained in the main text.

\section{Parameter estimation procedure based on linear programming}\label{app:bounds_LP}

For completeness, in this Appendix we summarize the parameter estimation technique presented in~\cite{Lo_Ma}, using linear programming, to evaluate the case of perfect discrete phase randomization for the protocol described in Sec.~\ref{Protocol_Key}.

In particular, given that the PDF  follows Eq.~\eqref{eq:dis_perfect}, which we will denote as $f(\theta)$ as in the previous Appendix and $N \geq 1$, a purification of Alice's emitted states can be expressed as
\begin{equation}
\begin{aligned}\label{mon_e}
|\psi^{\mu,N}_{[f(\theta)]}\rangle &=\sum_{k=0}^{N-1}|k\rangle_A| \sqrt{\mu} e^{2 k \pi i / N}\rangle \\
&=\sum_{j=0}^{N-1}|j\rangle_A\left|\beta^{\mu}_j\right\rangle,
\end{aligned}
\end{equation}
where the second equality corresponds to the Schmidt decomposition. Note that in Eq.~(\ref{mon_e}) we consider unnormalized states, which we will do throughout this Appendix for convenience. The states $\ket{j}_{A}$ can be interpreted as a quantum coin with $N$ random outputs, while the states $|\beta^{\mu}_j\rangle$ are given by
\begin{equation}
|\beta^{\mu}_j\rangle=\sum_{k=0}^{N-1} e^{-2 k j \pi i / N}|e^{2 k \pi i / N} \sqrt{\mu}\rangle.
\end{equation}
By using Eq.~\eqref{eq:coherent}, these latter states can be rewritten as
\begin{equation}
\left|\beta^{\mu}_j\right\rangle=\sum_{l=0}^{\infty} \frac{(\sqrt{\mu})^{l N+j}}{\sqrt{(l N+j) !}}|l N+j\rangle.
\end{equation}
Indeed, it is easy to show that when $N$ is large, $|\beta^{\mu}_j\rangle$ approaches a Fock state with $j$ photons.  

If Alice measures her ancilla system $A$ from the state $|\psi^{\mu,N}_{[f(\theta)]}\rangle$ in the basis $\{\ket{j}_A\}$, she obtains the result $j$ with probability $P^{\mu}_j$ given by
\begin{eqnarray}
\label{eq:pns_Lo}
P^{\mu}_j &=& \frac{\left\langle\beta^{\mu}_j \mid \beta^{\mu}_j\right\rangle}{\sum_{j=0}^{N-1}\left\langle\beta^{\mu}_j \mid \beta^{\mu}_j\right\rangle} \\
&=&\sum_{l=0}^{\infty} \frac{\mu^{l N+j} e^{-\mu}}{(l N+j) !}.
\end{eqnarray}

Ref.~\cite{Lo_Ma} employs the GLLP security analysis~\cite{GLLP}, which needs to determine the basis dependence $\Delta^{\mu}_{j}$ of the source, which is closely related to the fidelity $F^{\mu}_j$ between the states in the $X$ and $Z$ basis. Precisely, let us define 
\begin{equation}
\label{eq:basis_dependence}
 \Delta^{\mu}_{j}=\frac{1-F^{\mu}_j}{2 Y^{Z}_{j,\mu, f(\theta)}}, 
 \end{equation}
where $Y^{Z}_{j,\mu, f(\theta)}$ refers to the yield that corresponds to the states $|\beta^{\mu}_j\rangle$ encoded in the Z basis, and the fidelity $F^{\mu}_j$ can be bounded by
\begin{equation}
\begin{aligned}
 &F^{\mu}_j\geq\left|\frac{\sum_{l=0}^{\infty} \frac{\mu^{l N+j}}{(l N+j) !} 2^{-\frac{l N+j}{2}}\left(\cos \frac{l N+j}{4} \pi+\sin \frac{l N+j}{4} \pi\right)}{\sum_{l=0}^{\infty} \frac{\mu^{l N+j}}{(l N+j) !}}\right|.
\end{aligned}
\end{equation}

Moreover, since $\left|\beta_j^{\mu}\right\rangle \neq\left|\beta_j^{\gamma}\right\rangle$ when $\mu\neq\gamma$, one can relate the yields and bit error rates associated to different intensity settings as follows~\cite{GLLP}
\begin{equation}
\begin{aligned}
&\left|Y_{j,\mu, f(\theta)}-Y_{j,\gamma, f(\theta)}\right| \leq \sqrt{1-F_{\mu \gamma}^2}, \\
&\left|e^b_{j,\mu, f(\theta)} Y_{j,\mu, f(\theta)}-e^b_{j,\gamma, f(\theta)} Y_{j,\gamma, f(\theta)}\right| \leq \sqrt{1-F_{\mu \gamma}^2},
\end{aligned}
\end{equation}
where $e^b_{j,\mu, f(\theta)}$ denotes the bit error rate corresponding to the states $|\beta_j^{\mu}\rangle$, {\it i.e.}, the probability that Alice and Bob obtain different results when they use the same basis and Alice emits the state $|\beta_j^{\mu}\rangle$. The parameter $F_{\mu\gamma}$, on the other hand, is given by 
\begin{equation}
F_{\mu\gamma} := \frac{\sum_{l=0}^{\infty} \frac{(\mu \gamma)^{l N / 2}}{(l N) !}}{\sqrt{\sum_{l=0}^{\infty} \frac{\mu^{l N}}{(l N) !} \sum_{l=0}^{\infty} \frac{\gamma^{l N}}{(l N) !}}}.
\end{equation}

The phase error rate $e_{j,\mu,g(\theta)}$ in the $Z$ basis can be upper bounded by means of the bit error rate $e_{j,\mu, g(\theta)}^b$ in the $X$ basis and the basis dependence parameter $ \Delta^{\mu}_{j}$ as \cite{Lo_Preskill}
\begin{eqnarray}\label{qwe}
&e_{j,\mu,f(\theta)} \leq e_{j,\mu, f(\theta)}^{b, X}+4 \Delta^{\mu}_{j}\left(1-\Delta^{\mu}_{j}\right)\left(1-2 e^{b, X}_{j,\mu, f(\theta)}\right)\nonumber \\
&+4\left(1-2 \Delta^{\mu}_{j}\right) \sqrt{\Delta^{\mu}_{j}\left(1-\Delta^{\mu}_{j}\right) e^{b, X}_{j,\mu, f(\theta)}\left(1-e^{b, X}_{j,\mu, f(\theta)}\right)},\nonumber \\
\end{eqnarray}
where we have included the superscript $X$ in the bit error rate to emphasize that it refers to that in the $X$ basis. 

Putting it all together, we have that a lower bound on the yields $Y^Z_{j,s, f(\theta)}$ encoded in the Z basis can be estimated with the following linear program
\begin{gather}\label{LP_final}
\begin{aligned} 
\textup{min}\hspace{.1cm}& Y^Z_{j,s, f(\theta)} \\
\textup{s.t.}\hspace{.1cm}&\left|Y^Z_{j,\mu, f(\theta)}-Y^Z_{j,\gamma, f(\theta)}\right| \leq \sqrt{1-F_{\mu \gamma}^2},  \\
& \forall \mu, \gamma \in \{s, \nu, \omega \},\ \mu\neq\gamma, \\
& Q_{\mu, f(\theta)}^Z=\sum_{j=0}^{N-1} P_j^\mu Y^Z_{j,\mu, f(\theta)}, \forall \mu \in \{s, \nu, \omega \}.
\end{aligned}
\end{gather}

Similarly, an upper bound on the bit error rate $e_{j,\mu, g(\theta)}^{b, X}$ can be calculated with the following linear program
\begin{gather}\label{LP_final_error}
\begin{aligned} 
\textup{max}\hspace{.1cm}& \xi^X_{j,s, f(\theta)} \\
\textup{s.t.}\hspace{.1cm}&\left|\xi^X_{j,\mu, f(\theta)}-\xi^X_{j,\gamma, f(\theta)} \right| \leq \sqrt{1-F_{\mu \gamma}^2}, \\
& \forall \mu, \gamma \in \{s, \nu, \omega \},\ \mu\neq\gamma, \\
& E^X_{\mu, f(\theta)} Q^X_{\mu, f(\theta)}=\sum_{j=0}^{N-1} P_j^\mu \xi^X_{j,\mu, f(\theta)}, \forall \mu \in \{s, \nu, \omega \},
\end{aligned}
\end{gather}
where $\xi^{X}_{j,s, f(\theta)}=e^{b, X}_{j,s, f(\theta)} Y^X_{j,s, f(\theta)}$. In particular, let $\xi^{X*}_{j,s, f(\theta)}$ denote the solution to the linear program above, then we have that
\begin{equation}
e^{b, X}_{j,s, f(\theta)}\leq\frac{\xi^{X*}_{j,s, f(\theta)}}{Y^{X, {\rm L}}_{j,s, f(\theta)}}:=e^{b, X, {\rm U}}_{j,s, f(\theta)},
\end{equation}
where $Y^{X, {\rm L}}_{j,s, f(\theta)}$ represents a lower bound on the yield $Y^{X}_{j,s, f(\theta)}$ in the $X$ basis. This quantity can be calculated with the linear program given by Eq.~(\ref{LP_final}) by simply replacing the superscript $Z$ with $X$. 

Finally, one can calculate the phase error rate $e_{j,\mu,f(\theta)}$ in the $Z$ basis by means of Eq.~(\ref{qwe}), after replacing $e_{j,\mu, f(\theta)}^{b, X}$ with its upper bound and $ \Delta^{\mu}_{j}$ with the upper bound obtained after replacing a lower bound for the yield in Eq.~\eqref{eq:basis_dependence}. Importantly, with this approach there is no need to make a projection onto a finite dimensional subspace. This means that when evaluating the secret key rate formula given by Eq.~\eqref{eq:skr_general}, the probabilities $p^{L}_{n\mid s, f(\theta)}$ are directly given by $P^{\mu}_j$ as defined in Eq.~\eqref{eq:pns_Lo}. 

\nocite{*}

\bibliography{paper}

\providecommand{\noopsort}[1]{}\providecommand{\singleletter}[1]{#1}%
\begin{thebibliography}{58}%
\makeatletter
\providecommand \@ifxundefined [1]{%
 \@ifx{#1\undefined}
}%
\providecommand \@ifnum [1]{%
 \ifnum #1\expandafter \@firstoftwo
 \else \expandafter \@secondoftwo
 \fi
}%
\providecommand \@ifx [1]{%
 \ifx #1\expandafter \@firstoftwo
 \else \expandafter \@secondoftwo
 \fi
}%
\providecommand \natexlab [1]{#1}%
\providecommand \enquote  [1]{``#1''}%
\providecommand \bibnamefont  [1]{#1}%
\providecommand \bibfnamefont [1]{#1}%
\providecommand \citenamefont [1]{#1}%
\providecommand \href@noop [0]{\@secondoftwo}%
\providecommand \href [0]{\begingroup \@sanitize@url \@href}%
\providecommand \@href[1]{\@@startlink{#1}\@@href}%
\providecommand \@@href[1]{\endgroup#1\@@endlink}%
\providecommand \@sanitize@url [0]{\catcode `\\12\catcode `\$12\catcode
  `\&12\catcode `\#12\catcode `\^12\catcode `\_12\catcode `\%12\relax}%
\providecommand \@@startlink[1]{}%
\providecommand \@@endlink[0]{}%
\providecommand \url  [0]{\begingroup\@sanitize@url \@url }%
\providecommand \@url [1]{\endgroup\@href {#1}{\urlprefix }}%
\providecommand \urlprefix  [0]{URL }%
\providecommand \Eprint [0]{\href }%
\providecommand \doibase [0]{https://doi.org/}%
\providecommand \selectlanguage [0]{\@gobble}%
\providecommand \bibinfo  [0]{\@secondoftwo}%
\providecommand \bibfield  [0]{\@secondoftwo}%
\providecommand \translation [1]{[#1]}%
\providecommand \BibitemOpen [0]{}%
\providecommand \bibitemStop [0]{}%
\providecommand \bibitemNoStop [0]{.\EOS\space}%
\providecommand \EOS [0]{\spacefactor3000\relax}%
\providecommand \BibitemShut  [1]{\csname bibitem#1\endcsname}%
\let\auto@bib@innerbib\@empty
\bibitem [{\citenamefont {Xu}\ \emph {et~al.}(2020)\citenamefont {Xu},
  \citenamefont {Ma}, \citenamefont {Zhang}, \citenamefont {Lo},\ and\
  \citenamefont {Pan}}]{extra1}%
  \BibitemOpen
  \bibfield  {author} {\bibinfo {author} {\bibfnamefont {F.}~\bibnamefont
  {Xu}}, \bibinfo {author} {\bibfnamefont {X.}~\bibnamefont {Ma}}, \bibinfo
  {author} {\bibfnamefont {Q.}~\bibnamefont {Zhang}}, \bibinfo {author}
  {\bibfnamefont {H.-K.}\ \bibnamefont {Lo}},\ and\ \bibinfo {author}
  {\bibfnamefont {J.-W.}\ \bibnamefont {Pan}},\ }\bibfield  {title} {\bibinfo
  {title} {Secure quantum key distribution with realistic devices},\ }\href
  {https://doi.org/10.1103/RevModPhys.92.025002} {\bibfield  {journal}
  {\bibinfo  {journal} {Reviews of Modern Physics}\ }\textbf {\bibinfo {volume}
  {92}},\ \bibinfo {pages} {025002} (\bibinfo {year} {2020})}\BibitemShut
  {NoStop}%
\bibitem [{\citenamefont {Pirandola}\ \emph {et~al.}(2020)\citenamefont
  {Pirandola}, \citenamefont {Andersen}, \citenamefont {Banchi}, \citenamefont
  {Berta}, \citenamefont {Bunandar}, \citenamefont {Colbeck}, \citenamefont
  {Englund}, \citenamefont {Gehring}, \citenamefont {Lupo}, \citenamefont
  {Ottaviani} \emph {et~al.}}]{extra2}%
  \BibitemOpen
  \bibfield  {author} {\bibinfo {author} {\bibfnamefont {S.}~\bibnamefont
  {Pirandola}}, \bibinfo {author} {\bibfnamefont {U.~L.}\ \bibnamefont
  {Andersen}}, \bibinfo {author} {\bibfnamefont {L.}~\bibnamefont {Banchi}},
  \bibinfo {author} {\bibfnamefont {M.}~\bibnamefont {Berta}}, \bibinfo
  {author} {\bibfnamefont {D.}~\bibnamefont {Bunandar}}, \bibinfo {author}
  {\bibfnamefont {R.}~\bibnamefont {Colbeck}}, \bibinfo {author} {\bibfnamefont
  {D.}~\bibnamefont {Englund}}, \bibinfo {author} {\bibfnamefont
  {T.}~\bibnamefont {Gehring}}, \bibinfo {author} {\bibfnamefont
  {C.}~\bibnamefont {Lupo}}, \bibinfo {author} {\bibfnamefont {C.}~\bibnamefont
  {Ottaviani}}, \emph {et~al.},\ }\bibfield  {title} {\bibinfo {title}
  {Advances in quantum cryptography},\ }\href
  {https://doi.org/10.1364/aop.361502} {\bibfield  {journal} {\bibinfo
  {journal} {Advances in Optics and Photonics}\ }\textbf {\bibinfo {volume}
  {12}},\ \bibinfo {pages} {1012} (\bibinfo {year} {2020})}\BibitemShut
  {NoStop}%
\bibitem [{\citenamefont {Lo}\ \emph {et~al.}(2014)\citenamefont {Lo},
  \citenamefont {Curty},\ and\ \citenamefont {Tamaki}}]{extra3}%
  \BibitemOpen
  \bibfield  {author} {\bibinfo {author} {\bibfnamefont {H.-K.}\ \bibnamefont
  {Lo}}, \bibinfo {author} {\bibfnamefont {M.}~\bibnamefont {Curty}},\ and\
  \bibinfo {author} {\bibfnamefont {K.}~\bibnamefont {Tamaki}},\ }\bibfield
  {title} {\bibinfo {title} {Secure quantum key distribution},\ }\href
  {https://doi.org/10.1038/nphoton.2014.149} {\bibfield  {journal} {\bibinfo
  {journal} {Nature Photonics}\ }\textbf {\bibinfo {volume} {8}},\ \bibinfo
  {pages} {595} (\bibinfo {year} {2014})}\BibitemShut {NoStop}%
\bibitem [{\citenamefont {Wootters}\ and\ \citenamefont
  {Zurek}(1982)}]{cloning}%
  \BibitemOpen
  \bibfield  {author} {\bibinfo {author} {\bibfnamefont {W.~K.}\ \bibnamefont
  {Wootters}}\ and\ \bibinfo {author} {\bibfnamefont {W.~H.}\ \bibnamefont
  {Zurek}},\ }\bibfield  {title} {\bibinfo {title} {A single quantum cannot be
  cloned},\ }\href {https://doi.org/10.1038/299802a0} {\bibfield  {journal}
  {\bibinfo  {journal} {Nature}\ }\textbf {\bibinfo {volume} {299}},\ \bibinfo
  {pages} {802–803} (\bibinfo {year} {1982})}\BibitemShut {NoStop}%
\bibitem [{\citenamefont {Vernam}(1926)}]{Vernam}%
  \BibitemOpen
  \bibfield  {author} {\bibinfo {author} {\bibfnamefont {G.~S.}\ \bibnamefont
  {Vernam}},\ }\bibfield  {title} {\bibinfo {title} {Cipher printing telegraph
  systems for secret wire and radio telegraphic communications},\ }\href
  {https://doi.org/10.1109/T-AIEE.1926.5061224} {\bibfield  {journal} {\bibinfo
   {journal} {Transactions of the American Institute of Electrical Engineers}\
  }\textbf {\bibinfo {volume} {XLV}},\ \bibinfo {pages} {295} (\bibinfo {year}
  {1926})}\BibitemShut {NoStop}%
\bibitem [{\citenamefont {Sasaki}\ \emph {et~al.}(2011)\citenamefont {Sasaki},
  \citenamefont {Fujiwara}, \citenamefont {Ishizuka}, \citenamefont {Klaus},
  \citenamefont {Wakui}, \citenamefont {Takeoka}, \citenamefont {Miki},
  \citenamefont {Yamashita}, \citenamefont {Wang}, \citenamefont {Tanaka} \emph
  {et~al.}}]{nueva1}%
  \BibitemOpen
  \bibfield  {author} {\bibinfo {author} {\bibfnamefont {M.}~\bibnamefont
  {Sasaki}}, \bibinfo {author} {\bibfnamefont {M.}~\bibnamefont {Fujiwara}},
  \bibinfo {author} {\bibfnamefont {H.}~\bibnamefont {Ishizuka}}, \bibinfo
  {author} {\bibfnamefont {W.}~\bibnamefont {Klaus}}, \bibinfo {author}
  {\bibfnamefont {K.}~\bibnamefont {Wakui}}, \bibinfo {author} {\bibfnamefont
  {M.}~\bibnamefont {Takeoka}}, \bibinfo {author} {\bibfnamefont
  {S.}~\bibnamefont {Miki}}, \bibinfo {author} {\bibfnamefont {T.}~\bibnamefont
  {Yamashita}}, \bibinfo {author} {\bibfnamefont {Z.}~\bibnamefont {Wang}},
  \bibinfo {author} {\bibfnamefont {A.}~\bibnamefont {Tanaka}}, \emph
  {et~al.},\ }\bibfield  {title} {\bibinfo {title} {Field test of quantum key
  distribution in the tokyo qkd network},\ }\href
  {https://doi.org/10.1364/oe.19.010387} {\bibfield  {journal} {\bibinfo
  {journal} {Optics Express}\ }\textbf {\bibinfo {volume} {19}},\ \bibinfo
  {pages} {10387} (\bibinfo {year} {2011})}\BibitemShut {NoStop}%
\bibitem [{\citenamefont {Stucki}\ \emph {et~al.}(2011)\citenamefont {Stucki},
  \citenamefont {Legr{\'{e}}}, \citenamefont {Buntschu}, \citenamefont
  {Clausen}, \citenamefont {Felber}, \citenamefont {Gisin}, \citenamefont
  {Henzen}, \citenamefont {Junod}, \citenamefont {Litzistorf}, \citenamefont
  {Monbaron} \emph {et~al.}}]{nueva2}%
  \BibitemOpen
  \bibfield  {author} {\bibinfo {author} {\bibfnamefont {D.}~\bibnamefont
  {Stucki}}, \bibinfo {author} {\bibfnamefont {M.}~\bibnamefont {Legr{\'{e}}}},
  \bibinfo {author} {\bibfnamefont {F.}~\bibnamefont {Buntschu}}, \bibinfo
  {author} {\bibfnamefont {B.}~\bibnamefont {Clausen}}, \bibinfo {author}
  {\bibfnamefont {N.}~\bibnamefont {Felber}}, \bibinfo {author} {\bibfnamefont
  {N.}~\bibnamefont {Gisin}}, \bibinfo {author} {\bibfnamefont
  {L.}~\bibnamefont {Henzen}}, \bibinfo {author} {\bibfnamefont
  {P.}~\bibnamefont {Junod}}, \bibinfo {author} {\bibfnamefont
  {G.}~\bibnamefont {Litzistorf}}, \bibinfo {author} {\bibfnamefont
  {P.}~\bibnamefont {Monbaron}}, \emph {et~al.},\ }\bibfield  {title} {\bibinfo
  {title} {Long-term performance of the swissquantum quantum key distribution
  network in a field environment},\ }\href
  {https://doi.org/10.1088/1367-2630/13/12/123001} {\bibfield  {journal}
  {\bibinfo  {journal} {New Journal of Physics}\ }\textbf {\bibinfo {volume}
  {13}},\ \bibinfo {pages} {123001} (\bibinfo {year} {2011})}\BibitemShut
  {NoStop}%
\bibitem [{\citenamefont {Dynes}\ \emph {et~al.}(2019)\citenamefont {Dynes},
  \citenamefont {Wonfor}, \citenamefont {Tam}, \citenamefont {Sharpe},
  \citenamefont {Takahashi}, \citenamefont {Lucamarini}, \citenamefont {Plews},
  \citenamefont {Yuan}, \citenamefont {Dixon}, \citenamefont {Cho} \emph
  {et~al.}}]{nueva3}%
  \BibitemOpen
  \bibfield  {author} {\bibinfo {author} {\bibfnamefont {J.~F.}\ \bibnamefont
  {Dynes}}, \bibinfo {author} {\bibfnamefont {A.}~\bibnamefont {Wonfor}},
  \bibinfo {author} {\bibfnamefont {W.~W.~S.}\ \bibnamefont {Tam}}, \bibinfo
  {author} {\bibfnamefont {A.~W.}\ \bibnamefont {Sharpe}}, \bibinfo {author}
  {\bibfnamefont {R.}~\bibnamefont {Takahashi}}, \bibinfo {author}
  {\bibfnamefont {M.}~\bibnamefont {Lucamarini}}, \bibinfo {author}
  {\bibfnamefont {A.}~\bibnamefont {Plews}}, \bibinfo {author} {\bibfnamefont
  {Z.~L.}\ \bibnamefont {Yuan}}, \bibinfo {author} {\bibfnamefont {A.~R.}\
  \bibnamefont {Dixon}}, \bibinfo {author} {\bibfnamefont {J.}~\bibnamefont
  {Cho}}, \emph {et~al.},\ }\bibfield  {title} {\bibinfo {title} {Cambridge
  quantum network},\ }\href {https://doi.org/10.1038/s41534-019-0221-4}
  {\bibfield  {journal} {\bibinfo  {journal} {npj Quantum Information}\
  }\textbf {\bibinfo {volume} {5}},\ \bibinfo {pages} {101} (\bibinfo {year}
  {2019})}\BibitemShut {NoStop}%
\bibitem [{\citenamefont {Chen}\ \emph {et~al.}(2021)\citenamefont {Chen},
  \citenamefont {Zhang}, \citenamefont {Chen}, \citenamefont {Cai},
  \citenamefont {Liao}, \citenamefont {Zhang}, \citenamefont {Chen},
  \citenamefont {Yin}, \citenamefont {Ren}, \citenamefont {Chen} \emph
  {et~al.}}]{nueva4}%
  \BibitemOpen
  \bibfield  {author} {\bibinfo {author} {\bibfnamefont {Y.-A.}\ \bibnamefont
  {Chen}}, \bibinfo {author} {\bibfnamefont {Q.}~\bibnamefont {Zhang}},
  \bibinfo {author} {\bibfnamefont {T.-Y.}\ \bibnamefont {Chen}}, \bibinfo
  {author} {\bibfnamefont {W.-Q.}\ \bibnamefont {Cai}}, \bibinfo {author}
  {\bibfnamefont {S.-K.}\ \bibnamefont {Liao}}, \bibinfo {author}
  {\bibfnamefont {J.}~\bibnamefont {Zhang}}, \bibinfo {author} {\bibfnamefont
  {K.}~\bibnamefont {Chen}}, \bibinfo {author} {\bibfnamefont {J.}~\bibnamefont
  {Yin}}, \bibinfo {author} {\bibfnamefont {J.-G.}\ \bibnamefont {Ren}},
  \bibinfo {author} {\bibfnamefont {Z.}~\bibnamefont {Chen}}, \emph {et~al.},\
  }\bibfield  {title} {\bibinfo {title} {An integrated space-to-ground quantum
  communication network over 4,600 kilometres},\ }\href
  {https://doi.org/10.1038/s41586-020-03093-8} {\bibfield  {journal} {\bibinfo
  {journal} {Nature}\ }\textbf {\bibinfo {volume} {589}},\ \bibinfo {pages}
  {214} (\bibinfo {year} {2021})}\BibitemShut {NoStop}%
\bibitem [{\citenamefont {Bennett}\ and\ \citenamefont
  {Brassard}(1984)}]{BB84}%
  \BibitemOpen
  \bibfield  {author} {\bibinfo {author} {\bibfnamefont {C.~H.}\ \bibnamefont
  {Bennett}}\ and\ \bibinfo {author} {\bibfnamefont {G.}~\bibnamefont
  {Brassard}},\ }\bibfield  {title} {\bibinfo {title} {Quantum cryptography:
  Public key distribution and coin tossing},\ }in\ \href@noop {} {\emph
  {\bibinfo {booktitle} {Proceedings of IEEE International Conference on
  Computers, Systems, and Signal Processing}}}\ (\bibinfo {year} {1984})\ pp.\
  \bibinfo {pages} {175--179}\BibitemShut {NoStop}%
\bibitem [{\citenamefont {Huttner}\ \emph {et~al.}(1995)\citenamefont
  {Huttner}, \citenamefont {Imoto}, \citenamefont {Gisin},\ and\ \citenamefont
  {Mor}}]{PNS1}%
  \BibitemOpen
  \bibfield  {author} {\bibinfo {author} {\bibfnamefont {B.}~\bibnamefont
  {Huttner}}, \bibinfo {author} {\bibfnamefont {N.}~\bibnamefont {Imoto}},
  \bibinfo {author} {\bibfnamefont {N.}~\bibnamefont {Gisin}},\ and\ \bibinfo
  {author} {\bibfnamefont {T.}~\bibnamefont {Mor}},\ }\bibfield  {title}
  {\bibinfo {title} {Quantum cryptography with coherent states},\ }\href
  {https://doi.org/10.1103/PhysRevA.51.1863} {\bibfield  {journal} {\bibinfo
  {journal} {Physical Review A}\ }\textbf {\bibinfo {volume} {51}},\ \bibinfo
  {pages} {1863} (\bibinfo {year} {1995})}\BibitemShut {NoStop}%
\bibitem [{\citenamefont {Brassard}\ \emph {et~al.}(2000)\citenamefont
  {Brassard}, \citenamefont {Lütkenhaus}, \citenamefont {Mor},\ and\
  \citenamefont {Sanders}}]{PNS2}%
  \BibitemOpen
  \bibfield  {author} {\bibinfo {author} {\bibfnamefont {G.}~\bibnamefont
  {Brassard}}, \bibinfo {author} {\bibfnamefont {N.}~\bibnamefont
  {Lütkenhaus}}, \bibinfo {author} {\bibfnamefont {T.}~\bibnamefont {Mor}},\
  and\ \bibinfo {author} {\bibfnamefont {B.~C.}\ \bibnamefont {Sanders}},\
  }\bibfield  {title} {\bibinfo {title} {Limitations on practical quantum
  cryptography},\ }\href {https://doi.org/10.1103/PhysRevLett.85.1330}
  {\bibfield  {journal} {\bibinfo  {journal} {Physical Review Letters}\
  }\textbf {\bibinfo {volume} {85}},\ \bibinfo {pages} {1330} (\bibinfo {year}
  {2000})}\BibitemShut {NoStop}%
\bibitem [{\citenamefont {Hwang}(2003)}]{decoy1}%
  \BibitemOpen
  \bibfield  {author} {\bibinfo {author} {\bibfnamefont {W.-Y.}\ \bibnamefont
  {Hwang}},\ }\bibfield  {title} {\bibinfo {title} {Quantum key distribution
  with high loss: Toward global secure communication},\ }\href
  {https://doi.org/10.1103/physrevlett.91.057901} {\bibfield  {journal}
  {\bibinfo  {journal} {Physical Review Letters}\ }\textbf {\bibinfo {volume}
  {91}},\ \bibinfo {pages} {057901} (\bibinfo {year} {2003})}\BibitemShut
  {NoStop}%
\bibitem [{\citenamefont {Wang}(2005)}]{decoy2}%
  \BibitemOpen
  \bibfield  {author} {\bibinfo {author} {\bibfnamefont {X.-B.}\ \bibnamefont
  {Wang}},\ }\bibfield  {title} {\bibinfo {title} {Beating the
  photon-number-splitting attack in practical quantum cryptography},\ }\href
  {https://doi.org/10.1103/physrevlett.94.230503} {\bibfield  {journal}
  {\bibinfo  {journal} {Physical Review Letters}\ }\textbf {\bibinfo {volume}
  {94}},\ \bibinfo {pages} {230503} (\bibinfo {year} {2005})}\BibitemShut
  {NoStop}%
\bibitem [{\citenamefont {Lo}\ \emph {et~al.}(2005)\citenamefont {Lo},
  \citenamefont {Ma},\ and\ \citenamefont {Chen}}]{decoy3}%
  \BibitemOpen
  \bibfield  {author} {\bibinfo {author} {\bibfnamefont {H.-K.}\ \bibnamefont
  {Lo}}, \bibinfo {author} {\bibfnamefont {X.}~\bibnamefont {Ma}},\ and\
  \bibinfo {author} {\bibfnamefont {K.}~\bibnamefont {Chen}},\ }\bibfield
  {title} {\bibinfo {title} {Decoy state quantum key distribution},\ }\href
  {https://doi.org/10.1103/physrevlett.94.230504} {\bibfield  {journal}
  {\bibinfo  {journal} {Physical Review Letters}\ }\textbf {\bibinfo {volume}
  {94}},\ \bibinfo {pages} {230504} (\bibinfo {year} {2005})}\BibitemShut
  {NoStop}%
\bibitem [{\citenamefont {Lim}\ \emph {et~al.}(2014)\citenamefont {Lim},
  \citenamefont {Curty}, \citenamefont {Walenta}, \citenamefont {Xu},\ and\
  \citenamefont {Zbinden}}]{linear}%
  \BibitemOpen
  \bibfield  {author} {\bibinfo {author} {\bibfnamefont {C.~C.~W.}\
  \bibnamefont {Lim}}, \bibinfo {author} {\bibfnamefont {M.}~\bibnamefont
  {Curty}}, \bibinfo {author} {\bibfnamefont {N.}~\bibnamefont {Walenta}},
  \bibinfo {author} {\bibfnamefont {F.}~\bibnamefont {Xu}},\ and\ \bibinfo
  {author} {\bibfnamefont {H.}~\bibnamefont {Zbinden}},\ }\bibfield  {title}
  {\bibinfo {title} {Concise security bounds for practical decoy-state quantum
  key distribution},\ }\href {https://doi.org/10.1103/physreva.89.022307}
  {\bibfield  {journal} {\bibinfo  {journal} {Physical Review A}\ }\textbf
  {\bibinfo {volume} {89}},\ \bibinfo {pages} {022307} (\bibinfo {year}
  {2014})}\BibitemShut {NoStop}%
\bibitem [{\citenamefont {Zhao}\ \emph {et~al.}(2006)\citenamefont {Zhao},
  \citenamefont {Qi}, \citenamefont {Ma}, \citenamefont {Lo},\ and\
  \citenamefont {Qian}}]{demos1}%
  \BibitemOpen
  \bibfield  {author} {\bibinfo {author} {\bibfnamefont {Y.}~\bibnamefont
  {Zhao}}, \bibinfo {author} {\bibfnamefont {B.}~\bibnamefont {Qi}}, \bibinfo
  {author} {\bibfnamefont {X.}~\bibnamefont {Ma}}, \bibinfo {author}
  {\bibfnamefont {H.-K.}\ \bibnamefont {Lo}},\ and\ \bibinfo {author}
  {\bibfnamefont {L.}~\bibnamefont {Qian}},\ }\bibfield  {title} {\bibinfo
  {title} {Experimental quantum key distribution with decoy states},\ }\href
  {https://doi.org/10.1103/PhysRevLett.96.070502} {\bibfield  {journal}
  {\bibinfo  {journal} {Physical Review Letters}\ }\textbf {\bibinfo {volume}
  {96}},\ \bibinfo {pages} {70502} (\bibinfo {year} {2006})}\BibitemShut
  {NoStop}%
\bibitem [{\citenamefont {Rosenberg}\ \emph {et~al.}(2007)\citenamefont
  {Rosenberg}, \citenamefont {Harrington}, \citenamefont {Rice}, \citenamefont
  {Hiskett}, \citenamefont {Peterson}, \citenamefont {Hughes}, \citenamefont
  {Lita}, \citenamefont {Nam},\ and\ \citenamefont {Nordholt}}]{demos2}%
  \BibitemOpen
  \bibfield  {author} {\bibinfo {author} {\bibfnamefont {D.}~\bibnamefont
  {Rosenberg}}, \bibinfo {author} {\bibfnamefont {J.~W.}\ \bibnamefont
  {Harrington}}, \bibinfo {author} {\bibfnamefont {P.~R.}\ \bibnamefont
  {Rice}}, \bibinfo {author} {\bibfnamefont {P.~A.}\ \bibnamefont {Hiskett}},
  \bibinfo {author} {\bibfnamefont {C.~G.}\ \bibnamefont {Peterson}}, \bibinfo
  {author} {\bibfnamefont {R.~J.}\ \bibnamefont {Hughes}}, \bibinfo {author}
  {\bibfnamefont {A.~E.}\ \bibnamefont {Lita}}, \bibinfo {author}
  {\bibfnamefont {S.~W.}\ \bibnamefont {Nam}},\ and\ \bibinfo {author}
  {\bibfnamefont {J.~E.}\ \bibnamefont {Nordholt}},\ }\bibfield  {title}
  {\bibinfo {title} {Long-distance decoy-state quantum key distribution in
  optical fiber},\ }\href {https://doi.org/10.1103/physrevlett.98.010503}
  {\bibfield  {journal} {\bibinfo  {journal} {Physical Review Letters}\
  }\textbf {\bibinfo {volume} {98}},\ \bibinfo {pages} {10503} (\bibinfo {year}
  {2007})}\BibitemShut {NoStop}%
\bibitem [{\citenamefont {Schmitt-Manderbach}\ \emph
  {et~al.}(2007)\citenamefont {Schmitt-Manderbach}, \citenamefont {Weier},
  \citenamefont {F\"urst}, \citenamefont {Ursin}, \citenamefont {Tiefenbacher},
  \citenamefont {Scheidl}, \citenamefont {Perdigues}, \citenamefont {Sodnik},
  \citenamefont {Kurtsiefer}, \citenamefont {Rarity} \emph {et~al.}}]{demos3}%
  \BibitemOpen
  \bibfield  {author} {\bibinfo {author} {\bibfnamefont {T.}~\bibnamefont
  {Schmitt-Manderbach}}, \bibinfo {author} {\bibfnamefont {H.}~\bibnamefont
  {Weier}}, \bibinfo {author} {\bibfnamefont {M.}~\bibnamefont {F\"urst}},
  \bibinfo {author} {\bibfnamefont {R.}~\bibnamefont {Ursin}}, \bibinfo
  {author} {\bibfnamefont {F.}~\bibnamefont {Tiefenbacher}}, \bibinfo {author}
  {\bibfnamefont {T.}~\bibnamefont {Scheidl}}, \bibinfo {author} {\bibfnamefont
  {J.}~\bibnamefont {Perdigues}}, \bibinfo {author} {\bibfnamefont
  {Z.}~\bibnamefont {Sodnik}}, \bibinfo {author} {\bibfnamefont
  {C.}~\bibnamefont {Kurtsiefer}}, \bibinfo {author} {\bibfnamefont {J.~G.}\
  \bibnamefont {Rarity}}, \emph {et~al.},\ }\bibfield  {title} {\bibinfo
  {title} {Experimental demonstration of free-space decoy-state quantum key
  distribution over 144 km},\ }\href
  {https://doi.org/10.1103/PhysRevLett.98.010504} {\bibfield  {journal}
  {\bibinfo  {journal} {Physical Review Letters}\ }\textbf {\bibinfo {volume}
  {98}},\ \bibinfo {pages} {10504} (\bibinfo {year} {2007})}\BibitemShut
  {NoStop}%
\bibitem [{\citenamefont {Liu}\ \emph {et~al.}(2010{\natexlab{a}})\citenamefont
  {Liu}, \citenamefont {Chen}, \citenamefont {Wang}, \citenamefont {Cai},
  \citenamefont {Wan}, \citenamefont {Chen}, \citenamefont {Wang},
  \citenamefont {Liu}, \citenamefont {Liang}, \citenamefont {Yang} \emph
  {et~al.}}]{demos4}%
  \BibitemOpen
  \bibfield  {author} {\bibinfo {author} {\bibfnamefont {Y.}~\bibnamefont
  {Liu}}, \bibinfo {author} {\bibfnamefont {T.-Y.}\ \bibnamefont {Chen}},
  \bibinfo {author} {\bibfnamefont {J.}~\bibnamefont {Wang}}, \bibinfo {author}
  {\bibfnamefont {W.-Q.}\ \bibnamefont {Cai}}, \bibinfo {author} {\bibfnamefont
  {X.}~\bibnamefont {Wan}}, \bibinfo {author} {\bibfnamefont {L.-K.}\
  \bibnamefont {Chen}}, \bibinfo {author} {\bibfnamefont {J.-H.}\ \bibnamefont
  {Wang}}, \bibinfo {author} {\bibfnamefont {S.-B.}\ \bibnamefont {Liu}},
  \bibinfo {author} {\bibfnamefont {H.}~\bibnamefont {Liang}}, \bibinfo
  {author} {\bibfnamefont {L.}~\bibnamefont {Yang}}, \emph {et~al.},\
  }\bibfield  {title} {\bibinfo {title} {Decoy-state quantum key distribution
  with polarized photons over 200 km},\ }\href
  {https://doi.org/10.1364/OE.18.008587} {\bibfield  {journal} {\bibinfo
  {journal} {Optics Express}\ }\textbf {\bibinfo {volume} {18}},\ \bibinfo
  {pages} {8587} (\bibinfo {year} {2010}{\natexlab{a}})}\BibitemShut {NoStop}%
\bibitem [{\citenamefont {Fr\"{o}hlich}\ \emph {et~al.}(2017)\citenamefont
  {Fr\"{o}hlich}, \citenamefont {Lucamarini}, \citenamefont {Dynes},
  \citenamefont {Comandar}, \citenamefont {Tam}, \citenamefont {Plews},
  \citenamefont {Sharpe}, \citenamefont {Yuan},\ and\ \citenamefont
  {Shields}}]{demos5}%
  \BibitemOpen
  \bibfield  {author} {\bibinfo {author} {\bibfnamefont {B.}~\bibnamefont
  {Fr\"{o}hlich}}, \bibinfo {author} {\bibfnamefont {M.}~\bibnamefont
  {Lucamarini}}, \bibinfo {author} {\bibfnamefont {J.~F.}\ \bibnamefont
  {Dynes}}, \bibinfo {author} {\bibfnamefont {L.~C.}\ \bibnamefont {Comandar}},
  \bibinfo {author} {\bibfnamefont {W.~W.-S.}\ \bibnamefont {Tam}}, \bibinfo
  {author} {\bibfnamefont {A.}~\bibnamefont {Plews}}, \bibinfo {author}
  {\bibfnamefont {A.~W.}\ \bibnamefont {Sharpe}}, \bibinfo {author}
  {\bibfnamefont {Z.}~\bibnamefont {Yuan}},\ and\ \bibinfo {author}
  {\bibfnamefont {A.~J.}\ \bibnamefont {Shields}},\ }\bibfield  {title}
  {\bibinfo {title} {Long-distance quantum key distribution secure against
  coherent attacks},\ }\href {https://doi.org/10.1364/OPTICA.4.000163}
  {\bibfield  {journal} {\bibinfo  {journal} {Optica}\ }\textbf {\bibinfo
  {volume} {4}},\ \bibinfo {pages} {163} (\bibinfo {year} {2017})}\BibitemShut
  {NoStop}%
\bibitem [{\citenamefont {Yuan}\ \emph
  {et~al.}(2018{\natexlab{a}})\citenamefont {Yuan}, \citenamefont {Murakami},
  \citenamefont {Kujiraoka}, \citenamefont {Lucamarini}, \citenamefont
  {Tanizawa}, \citenamefont {Sato}, \citenamefont {Shields}, \citenamefont
  {Plews}, \citenamefont {Takahashi}, \citenamefont {Doi} \emph
  {et~al.}}]{demos6}%
  \BibitemOpen
  \bibfield  {author} {\bibinfo {author} {\bibfnamefont {Z.}~\bibnamefont
  {Yuan}}, \bibinfo {author} {\bibfnamefont {A.}~\bibnamefont {Murakami}},
  \bibinfo {author} {\bibfnamefont {M.}~\bibnamefont {Kujiraoka}}, \bibinfo
  {author} {\bibfnamefont {M.}~\bibnamefont {Lucamarini}}, \bibinfo {author}
  {\bibfnamefont {Y.}~\bibnamefont {Tanizawa}}, \bibinfo {author}
  {\bibfnamefont {H.}~\bibnamefont {Sato}}, \bibinfo {author} {\bibfnamefont
  {A.~J.}\ \bibnamefont {Shields}}, \bibinfo {author} {\bibfnamefont
  {A.}~\bibnamefont {Plews}}, \bibinfo {author} {\bibfnamefont
  {R.}~\bibnamefont {Takahashi}}, \bibinfo {author} {\bibfnamefont
  {K.}~\bibnamefont {Doi}}, \emph {et~al.},\ }\bibfield  {title} {\bibinfo
  {title} {10-mb/s quantum key distribution},\ }\href
  {https://doi.org/10.1109/jlt.2018.2843136} {\bibfield  {journal} {\bibinfo
  {journal} {Journal of Lightwave Technology}\ }\textbf {\bibinfo {volume}
  {36}},\ \bibinfo {pages} {3427} (\bibinfo {year}
  {2018}{\natexlab{a}})}\BibitemShut {NoStop}%
\bibitem [{\citenamefont {Boaron}\ \emph {et~al.}(2018)\citenamefont {Boaron},
  \citenamefont {Boso}, \citenamefont {Rusca}, \citenamefont {Vulliez},
  \citenamefont {Autebert}, \citenamefont {Caloz}, \citenamefont {Perrenoud},
  \citenamefont {Gras}, \citenamefont {Bussières}, \citenamefont {Li} \emph
  {et~al.}}]{gain_switch_5}%
  \BibitemOpen
  \bibfield  {author} {\bibinfo {author} {\bibfnamefont {A.}~\bibnamefont
  {Boaron}}, \bibinfo {author} {\bibfnamefont {G.}~\bibnamefont {Boso}},
  \bibinfo {author} {\bibfnamefont {D.}~\bibnamefont {Rusca}}, \bibinfo
  {author} {\bibfnamefont {C.}~\bibnamefont {Vulliez}}, \bibinfo {author}
  {\bibfnamefont {C.}~\bibnamefont {Autebert}}, \bibinfo {author}
  {\bibfnamefont {M.}~\bibnamefont {Caloz}}, \bibinfo {author} {\bibfnamefont
  {M.}~\bibnamefont {Perrenoud}}, \bibinfo {author} {\bibfnamefont
  {G.}~\bibnamefont {Gras}}, \bibinfo {author} {\bibfnamefont {F.}~\bibnamefont
  {Bussières}}, \bibinfo {author} {\bibfnamefont {M.-J.}\ \bibnamefont {Li}},
  \emph {et~al.},\ }\bibfield  {title} {\bibinfo {title} {Secure quantum key
  distribution over 421 km of optical fiber},\ }\href
  {https://doi.org/10.1103/PhysRevLett.121.190502} {\bibfield  {journal}
  {\bibinfo  {journal} {Phys. Rev. Lett.}\ }\textbf {\bibinfo {volume} {121}},\
  \bibinfo {pages} {190502} (\bibinfo {year} {2018})}\BibitemShut {NoStop}%
\bibitem [{\citenamefont {Liao}\ \emph {et~al.}(2017)\citenamefont {Liao},
  \citenamefont {Cai}, \citenamefont {Liu}, \citenamefont {Zhang},
  \citenamefont {Li}, \citenamefont {Ren}, \citenamefont {Yin}, \citenamefont
  {Shen}, \citenamefont {Cao}, \citenamefont {Li} \emph {et~al.}}]{sat1}%
  \BibitemOpen
  \bibfield  {author} {\bibinfo {author} {\bibfnamefont {S.-K.}\ \bibnamefont
  {Liao}}, \bibinfo {author} {\bibfnamefont {W.-Q.}\ \bibnamefont {Cai}},
  \bibinfo {author} {\bibfnamefont {W.-Y.}\ \bibnamefont {Liu}}, \bibinfo
  {author} {\bibfnamefont {L.}~\bibnamefont {Zhang}}, \bibinfo {author}
  {\bibfnamefont {Y.}~\bibnamefont {Li}}, \bibinfo {author} {\bibfnamefont
  {J.-G.}\ \bibnamefont {Ren}}, \bibinfo {author} {\bibfnamefont
  {J.}~\bibnamefont {Yin}}, \bibinfo {author} {\bibfnamefont {Q.}~\bibnamefont
  {Shen}}, \bibinfo {author} {\bibfnamefont {Y.}~\bibnamefont {Cao}}, \bibinfo
  {author} {\bibfnamefont {Z.-P.}\ \bibnamefont {Li}}, \emph {et~al.},\
  }\bibfield  {title} {\bibinfo {title} {Satellite-to-ground quantum key
  distribution},\ }\href {https://doi.org/10.1038/nature23655} {\bibfield
  {journal} {\bibinfo  {journal} {Nature}\ }\textbf {\bibinfo {volume} {549}},\
  \bibinfo {pages} {43} (\bibinfo {year} {2017})}\BibitemShut {NoStop}%
\bibitem [{\citenamefont {Liao}\ \emph {et~al.}(2018)\citenamefont {Liao},
  \citenamefont {Cai}, \citenamefont {Handsteiner}, \citenamefont {Liu},
  \citenamefont {Yin}, \citenamefont {Zhang}, \citenamefont {Rauch},
  \citenamefont {Fink}, \citenamefont {Ren}, \citenamefont {Liu} \emph
  {et~al.}}]{sat2}%
  \BibitemOpen
  \bibfield  {author} {\bibinfo {author} {\bibfnamefont {S.-K.}\ \bibnamefont
  {Liao}}, \bibinfo {author} {\bibfnamefont {W.-Q.}\ \bibnamefont {Cai}},
  \bibinfo {author} {\bibfnamefont {J.}~\bibnamefont {Handsteiner}}, \bibinfo
  {author} {\bibfnamefont {B.}~\bibnamefont {Liu}}, \bibinfo {author}
  {\bibfnamefont {J.}~\bibnamefont {Yin}}, \bibinfo {author} {\bibfnamefont
  {L.}~\bibnamefont {Zhang}}, \bibinfo {author} {\bibfnamefont
  {D.}~\bibnamefont {Rauch}}, \bibinfo {author} {\bibfnamefont
  {M.}~\bibnamefont {Fink}}, \bibinfo {author} {\bibfnamefont {J.-G.}\
  \bibnamefont {Ren}}, \bibinfo {author} {\bibfnamefont {W.-Y.}\ \bibnamefont
  {Liu}}, \emph {et~al.},\ }\bibfield  {title} {\bibinfo {title}
  {Satellite-relayed intercontinental quantum network},\ }\href
  {https://doi.org/10.1103/PhysRevLett.120.030501} {\bibfield  {journal}
  {\bibinfo  {journal} {Physical Review Letters}\ }\textbf {\bibinfo {volume}
  {120}},\ \bibinfo {pages} {030501} (\bibinfo {year} {2018})}\BibitemShut
  {NoStop}%
\bibitem [{\citenamefont {Sibson}\ \emph
  {et~al.}(2017{\natexlab{a}})\citenamefont {Sibson}, \citenamefont {Erven},
  \citenamefont {Godfrey}, \citenamefont {Miki}, \citenamefont {Yamashita},
  \citenamefont {Fujiwara}, \citenamefont {Sasaki}, \citenamefont {Terai},
  \citenamefont {Tanner}, \citenamefont {Natarajan} \emph {et~al.}}]{chip1}%
  \BibitemOpen
  \bibfield  {author} {\bibinfo {author} {\bibfnamefont {P.}~\bibnamefont
  {Sibson}}, \bibinfo {author} {\bibfnamefont {C.}~\bibnamefont {Erven}},
  \bibinfo {author} {\bibfnamefont {M.}~\bibnamefont {Godfrey}}, \bibinfo
  {author} {\bibfnamefont {S.}~\bibnamefont {Miki}}, \bibinfo {author}
  {\bibfnamefont {T.}~\bibnamefont {Yamashita}}, \bibinfo {author}
  {\bibfnamefont {M.}~\bibnamefont {Fujiwara}}, \bibinfo {author}
  {\bibfnamefont {M.}~\bibnamefont {Sasaki}}, \bibinfo {author} {\bibfnamefont
  {H.}~\bibnamefont {Terai}}, \bibinfo {author} {\bibfnamefont {M.~G.}\
  \bibnamefont {Tanner}}, \bibinfo {author} {\bibfnamefont {C.~M.}\
  \bibnamefont {Natarajan}}, \emph {et~al.},\ }\bibfield  {title} {\bibinfo
  {title} {Chip-based quantum key distribution},\ }\href
  {https://doi.org/10.1038/ncomms13984} {\bibfield  {journal} {\bibinfo
  {journal} {Nature Communications}\ }\textbf {\bibinfo {volume} {8}},\
  \bibinfo {pages} {13984} (\bibinfo {year} {2017}{\natexlab{a}})}\BibitemShut
  {NoStop}%
\bibitem [{\citenamefont {Bunandar}\ \emph
  {et~al.}(2018{\natexlab{a}})\citenamefont {Bunandar}, \citenamefont
  {Lentine}, \citenamefont {Lee}, \citenamefont {Cai}, \citenamefont {Long},
  \citenamefont {Boynton}, \citenamefont {Martinez}, \citenamefont {DeRose},
  \citenamefont {Chen}, \citenamefont {Grein} \emph {et~al.}}]{chip2}%
  \BibitemOpen
  \bibfield  {author} {\bibinfo {author} {\bibfnamefont {D.}~\bibnamefont
  {Bunandar}}, \bibinfo {author} {\bibfnamefont {A.}~\bibnamefont {Lentine}},
  \bibinfo {author} {\bibfnamefont {C.}~\bibnamefont {Lee}}, \bibinfo {author}
  {\bibfnamefont {H.}~\bibnamefont {Cai}}, \bibinfo {author} {\bibfnamefont
  {C.~M.}\ \bibnamefont {Long}}, \bibinfo {author} {\bibfnamefont
  {N.}~\bibnamefont {Boynton}}, \bibinfo {author} {\bibfnamefont
  {N.}~\bibnamefont {Martinez}}, \bibinfo {author} {\bibfnamefont
  {C.}~\bibnamefont {DeRose}}, \bibinfo {author} {\bibfnamefont
  {C.}~\bibnamefont {Chen}}, \bibinfo {author} {\bibfnamefont {M.}~\bibnamefont
  {Grein}}, \emph {et~al.},\ }\bibfield  {title} {\bibinfo {title}
  {Metropolitan quantum key distribution with silicon photonics},\ }\href
  {https://doi.org/10.1103/PhysRevX.8.021009} {\bibfield  {journal} {\bibinfo
  {journal} {Phys. Rev. X}\ }\textbf {\bibinfo {volume} {8}},\ \bibinfo {pages}
  {021009} (\bibinfo {year} {2018}{\natexlab{a}})}\BibitemShut {NoStop}%
\bibitem [{\citenamefont {Paraïso}\ \emph {et~al.}(2019)\citenamefont
  {Paraïso}, \citenamefont {De~Marco}, \citenamefont {Roger}, \citenamefont
  {Marangon}, \citenamefont {Dynes}, \citenamefont {Lucamarini}, \citenamefont
  {Yuan},\ and\ \citenamefont {Shields}}]{chip3}%
  \BibitemOpen
  \bibfield  {author} {\bibinfo {author} {\bibfnamefont {T.~K.}\ \bibnamefont
  {Paraïso}}, \bibinfo {author} {\bibfnamefont {I.}~\bibnamefont {De~Marco}},
  \bibinfo {author} {\bibfnamefont {T.}~\bibnamefont {Roger}}, \bibinfo
  {author} {\bibfnamefont {D.~G.}\ \bibnamefont {Marangon}}, \bibinfo {author}
  {\bibfnamefont {J.~F.}\ \bibnamefont {Dynes}}, \bibinfo {author}
  {\bibfnamefont {M.}~\bibnamefont {Lucamarini}}, \bibinfo {author}
  {\bibfnamefont {Z.}~\bibnamefont {Yuan}},\ and\ \bibinfo {author}
  {\bibfnamefont {A.~J.}\ \bibnamefont {Shields}},\ }\bibfield  {title}
  {\bibinfo {title} {A modulator-free quantum key distribution transmitter
  chip},\ }\href {https://doi.org/10.1038/s41534-019-0158-7} {\bibfield
  {journal} {\bibinfo  {journal} {npj Quantum Information}\ }\textbf {\bibinfo
  {volume} {5}},\ \bibinfo {pages} {42} (\bibinfo {year} {2019})}\BibitemShut
  {NoStop}%
\bibitem [{\citenamefont {Marco}\ \emph {et~al.}(2021)\citenamefont {Marco},
  \citenamefont {Woodward}, \citenamefont {Roberts}, \citenamefont
  {Para\"{i}so}, \citenamefont {Roger}, \citenamefont {Sanzaro}, \citenamefont
  {Lucamarini}, \citenamefont {Yuan},\ and\ \citenamefont {Shields}}]{chip4}%
  \BibitemOpen
  \bibfield  {author} {\bibinfo {author} {\bibfnamefont {I.~D.}\ \bibnamefont
  {Marco}}, \bibinfo {author} {\bibfnamefont {R.~I.}\ \bibnamefont {Woodward}},
  \bibinfo {author} {\bibfnamefont {G.~L.}\ \bibnamefont {Roberts}}, \bibinfo
  {author} {\bibfnamefont {T.~K.}\ \bibnamefont {Para\"{i}so}}, \bibinfo
  {author} {\bibfnamefont {T.}~\bibnamefont {Roger}}, \bibinfo {author}
  {\bibfnamefont {M.}~\bibnamefont {Sanzaro}}, \bibinfo {author} {\bibfnamefont
  {M.}~\bibnamefont {Lucamarini}}, \bibinfo {author} {\bibfnamefont
  {Z.}~\bibnamefont {Yuan}},\ and\ \bibinfo {author} {\bibfnamefont {A.~J.}\
  \bibnamefont {Shields}},\ }\bibfield  {title} {\bibinfo {title} {Real-time
  operation of a multi-rate, multi-protocol quantum key distribution
  transmitter},\ }\href {https://doi.org/10.1364/OPTICA.423517} {\bibfield
  {journal} {\bibinfo  {journal} {Optica}\ }\textbf {\bibinfo {volume} {8}},\
  \bibinfo {pages} {911} (\bibinfo {year} {2021})}\BibitemShut {NoStop}%
\bibitem [{IDQ()}]{IDQ}%
  \BibitemOpen
  \href@noop {} {\bibinfo {title} {Id quantique sa}},\ \bibinfo {howpublished}
  {\url{https://www.idquantique.com/}}\BibitemShut {NoStop}%
\bibitem [{Tos()}]{Toshiba}%
  \BibitemOpen
  \href@noop {} {\bibinfo {title} {Toshiba europe limited}},\ \bibinfo
  {howpublished}
  {\url{https://www.global.toshiba/ww/products-solutions/security-ict/qkd.html}}\BibitemShut
  {NoStop}%
\bibitem [{Qua()}]{QuantumCTek}%
  \BibitemOpen
  \href@noop {} {\bibinfo {title} {Quantumctek co., ltd.}},\ \bibinfo
  {howpublished} {\url{http://www.quantum-info.com/English/}}\BibitemShut
  {NoStop}%
\bibitem [{Thi()}]{ThinkQ}%
  \BibitemOpen
  \href@noop {} {\bibinfo {title} {Thinkquantum s.r.l.}},\ \bibinfo
  {howpublished} {\url{https://www.thinkquantum.com}}\BibitemShut {NoStop}%
\bibitem [{QTI()}]{QTI}%
  \BibitemOpen
  \href@noop {} {\bibinfo {title} {Quantum telecommunications italy s.r.l.}},\
  \bibinfo {howpublished} {\url{https://www.qticompany.com}}\BibitemShut
  {NoStop}%
\bibitem [{\citenamefont {Yuan}\ \emph {et~al.}(2007)\citenamefont {Yuan},
  \citenamefont {Sharpe},\ and\ \citenamefont {Shields}}]{gain_switch_1}%
  \BibitemOpen
  \bibfield  {author} {\bibinfo {author} {\bibfnamefont {Z.~L.}\ \bibnamefont
  {Yuan}}, \bibinfo {author} {\bibfnamefont {A.~W.}\ \bibnamefont {Sharpe}},\
  and\ \bibinfo {author} {\bibfnamefont {A.~J.}\ \bibnamefont {Shields}},\
  }\bibfield  {title} {\bibinfo {title} {Unconditionally secure one-way quantum
  key distribution using decoy pulses},\ }\href
  {https://doi.org/10.1063/1.2430685} {\bibfield  {journal} {\bibinfo
  {journal} {Appl. Phys. Lett.}\ }\textbf {\bibinfo {volume} {90}},\ \bibinfo
  {pages} {011118} (\bibinfo {year} {2007})}\BibitemShut {NoStop}%
\bibitem [{\citenamefont {Dixon}\ \emph {et~al.}(2008)\citenamefont {Dixon},
  \citenamefont {Yuan}, \citenamefont {Dynes}, \citenamefont {Sharpe},\ and\
  \citenamefont {Shields}}]{gain_switch_2}%
  \BibitemOpen
  \bibfield  {author} {\bibinfo {author} {\bibfnamefont {A.~R.}\ \bibnamefont
  {Dixon}}, \bibinfo {author} {\bibfnamefont {Z.~L.}\ \bibnamefont {Yuan}},
  \bibinfo {author} {\bibfnamefont {J.~F.}\ \bibnamefont {Dynes}}, \bibinfo
  {author} {\bibfnamefont {A.~W.}\ \bibnamefont {Sharpe}},\ and\ \bibinfo
  {author} {\bibfnamefont {A.~J.}\ \bibnamefont {Shields}},\ }\bibfield
  {title} {\bibinfo {title} {Gigahertz decoy quantum key distribution with 1
  mbit/s secure key rate},\ }\href {https://doi.org/10.1364/OE.16.018790}
  {\bibfield  {journal} {\bibinfo  {journal} {Opt. Express}\ }\textbf {\bibinfo
  {volume} {16}},\ \bibinfo {pages} {18790} (\bibinfo {year}
  {2008})}\BibitemShut {NoStop}%
\bibitem [{\citenamefont {Liu}\ \emph {et~al.}(2010{\natexlab{b}})\citenamefont
  {Liu}, \citenamefont {Chen}, \citenamefont {Wang}, \citenamefont {Cai},
  \citenamefont {Wan}, \citenamefont {Chen}, \citenamefont {Wang},
  \citenamefont {Liu}, \citenamefont {Liang}, \citenamefont {Yang} \emph
  {et~al.}}]{gain_switch_3}%
  \BibitemOpen
  \bibfield  {author} {\bibinfo {author} {\bibfnamefont {Y.}~\bibnamefont
  {Liu}}, \bibinfo {author} {\bibfnamefont {T.-Y.}\ \bibnamefont {Chen}},
  \bibinfo {author} {\bibfnamefont {J.}~\bibnamefont {Wang}}, \bibinfo {author}
  {\bibfnamefont {W.-Q.}\ \bibnamefont {Cai}}, \bibinfo {author} {\bibfnamefont
  {X.}~\bibnamefont {Wan}}, \bibinfo {author} {\bibfnamefont {L.-K.}\
  \bibnamefont {Chen}}, \bibinfo {author} {\bibfnamefont {J.-H.}\ \bibnamefont
  {Wang}}, \bibinfo {author} {\bibfnamefont {S.-B.}\ \bibnamefont {Liu}},
  \bibinfo {author} {\bibfnamefont {H.}~\bibnamefont {Liang}}, \bibinfo
  {author} {\bibfnamefont {L.}~\bibnamefont {Yang}}, \emph {et~al.},\
  }\bibfield  {title} {\bibinfo {title} {Decoy-state quantum key distribution
  with polarized photons over 200 km},\ }\href
  {https://doi.org/10.1364/OE.18.008587} {\bibfield  {journal} {\bibinfo
  {journal} {Opt. Express}\ }\textbf {\bibinfo {volume} {18}},\ \bibinfo
  {pages} {8587} (\bibinfo {year} {2010}{\natexlab{b}})}\BibitemShut {NoStop}%
\bibitem [{\citenamefont {Lucamarini}\ \emph {et~al.}(2013)\citenamefont
  {Lucamarini}, \citenamefont {Patel}, \citenamefont {Dynes}, \citenamefont
  {Fröhlich}, \citenamefont {Sharpe}, \citenamefont {Dixon}, \citenamefont
  {Yuan}, \citenamefont {Penty},\ and\ \citenamefont
  {Shields}}]{gain_switch_4}%
  \BibitemOpen
  \bibfield  {author} {\bibinfo {author} {\bibfnamefont {M.}~\bibnamefont
  {Lucamarini}}, \bibinfo {author} {\bibfnamefont {K.~A.}\ \bibnamefont
  {Patel}}, \bibinfo {author} {\bibfnamefont {J.~F.}\ \bibnamefont {Dynes}},
  \bibinfo {author} {\bibfnamefont {B.}~\bibnamefont {Fröhlich}}, \bibinfo
  {author} {\bibfnamefont {A.~W.}\ \bibnamefont {Sharpe}}, \bibinfo {author}
  {\bibfnamefont {A.~R.}\ \bibnamefont {Dixon}}, \bibinfo {author}
  {\bibfnamefont {Z.~L.}\ \bibnamefont {Yuan}}, \bibinfo {author}
  {\bibfnamefont {R.~V.}\ \bibnamefont {Penty}},\ and\ \bibinfo {author}
  {\bibfnamefont {A.~J.}\ \bibnamefont {Shields}},\ }\bibfield  {title}
  {\bibinfo {title} {Efficient decoy-state quantum key distribution with
  quantified security},\ }\href {https://doi.org/10.1364/oe.21.024550}
  {\bibfield  {journal} {\bibinfo  {journal} {Opt. Express}\ }\textbf {\bibinfo
  {volume} {21}},\ \bibinfo {pages} {21} (\bibinfo {year} {2013})}\BibitemShut
  {NoStop}%
\bibitem [{\citenamefont {Valivarthi}\ \emph {et~al.}(2017)\citenamefont
  {Valivarthi}, \citenamefont {Zhou}, \citenamefont {John}, \citenamefont
  {Marsili}, \citenamefont {Verma}, \citenamefont {Shaw}, \citenamefont {Nam},
  \citenamefont {Oblak},\ and\ \citenamefont {Tittel}}]{Valivarthi2017}%
  \BibitemOpen
  \bibfield  {author} {\bibinfo {author} {\bibfnamefont {R.}~\bibnamefont
  {Valivarthi}}, \bibinfo {author} {\bibfnamefont {Q.}~\bibnamefont {Zhou}},
  \bibinfo {author} {\bibfnamefont {C.}~\bibnamefont {John}}, \bibinfo {author}
  {\bibfnamefont {F.}~\bibnamefont {Marsili}}, \bibinfo {author} {\bibfnamefont
  {V.~B.}\ \bibnamefont {Verma}}, \bibinfo {author} {\bibfnamefont {M.~D.}\
  \bibnamefont {Shaw}}, \bibinfo {author} {\bibfnamefont {S.~W.}\ \bibnamefont
  {Nam}}, \bibinfo {author} {\bibfnamefont {D.}~\bibnamefont {Oblak}},\ and\
  \bibinfo {author} {\bibfnamefont {W.}~\bibnamefont {Tittel}},\ }\bibfield
  {title} {\bibinfo {title} {A cost-effective measurement-device-independent
  quantum key distribution system for quantum networks},\ }\href
  {https://doi.org/10.1088/2058-9565/aa8790} {\bibfield  {journal} {\bibinfo
  {journal} {Quantum Science and Technology}\ }\textbf {\bibinfo {volume}
  {2}},\ \bibinfo {pages} {04LT01} (\bibinfo {year} {2017})}\BibitemShut
  {NoStop}%
\bibitem [{\citenamefont {Yuan}\ \emph
  {et~al.}(2018{\natexlab{b}})\citenamefont {Yuan}, \citenamefont {Plews},
  \citenamefont {Takahashi}, \citenamefont {Doi}, \citenamefont {Tam},
  \citenamefont {Sharpe}, \citenamefont {Dixon}, \citenamefont {Lavelle},
  \citenamefont {Dynes}, \citenamefont {Murakami} \emph {et~al.}}]{Yuan2018}%
  \BibitemOpen
  \bibfield  {author} {\bibinfo {author} {\bibfnamefont {Z.}~\bibnamefont
  {Yuan}}, \bibinfo {author} {\bibfnamefont {A.}~\bibnamefont {Plews}},
  \bibinfo {author} {\bibfnamefont {R.}~\bibnamefont {Takahashi}}, \bibinfo
  {author} {\bibfnamefont {K.}~\bibnamefont {Doi}}, \bibinfo {author}
  {\bibfnamefont {W.}~\bibnamefont {Tam}}, \bibinfo {author} {\bibfnamefont
  {A.}~\bibnamefont {Sharpe}}, \bibinfo {author} {\bibfnamefont
  {A.}~\bibnamefont {Dixon}}, \bibinfo {author} {\bibfnamefont
  {E.}~\bibnamefont {Lavelle}}, \bibinfo {author} {\bibfnamefont
  {J.}~\bibnamefont {Dynes}}, \bibinfo {author} {\bibfnamefont
  {A.}~\bibnamefont {Murakami}}, \emph {et~al.},\ }\bibfield  {title} {\bibinfo
  {title} {10-mb/s quantum key distribution},\ }\href
  {https://doi.org/10.1109/jlt.2018.2843136} {\bibfield  {journal} {\bibinfo
  {journal} {Journal of Lightwave Technology}\ }\textbf {\bibinfo {volume}
  {36}},\ \bibinfo {pages} {3427} (\bibinfo {year}
  {2018}{\natexlab{b}})}\BibitemShut {NoStop}%
\bibitem [{\citenamefont {Zhao}\ \emph {et~al.}(2007)\citenamefont {Zhao},
  \citenamefont {Qi},\ and\ \citenamefont {Lo}}]{generation1}%
  \BibitemOpen
  \bibfield  {author} {\bibinfo {author} {\bibfnamefont {Y.}~\bibnamefont
  {Zhao}}, \bibinfo {author} {\bibfnamefont {B.}~\bibnamefont {Qi}},\ and\
  \bibinfo {author} {\bibfnamefont {H.-K.}\ \bibnamefont {Lo}},\ }\bibfield
  {title} {\bibinfo {title} {Experimental quantum key distribution with active
  phase randomization},\ }\href {https://doi.org/10.1063/1.2432296} {\bibfield
  {journal} {\bibinfo  {journal} {Applied Physics Letters}\ }\textbf {\bibinfo
  {volume} {90}},\ \bibinfo {pages} {044106} (\bibinfo {year}
  {2007})}\BibitemShut {NoStop}%
\bibitem [{\citenamefont {Sun}\ and\ \citenamefont
  {Liang}(2012)}]{generation2}%
  \BibitemOpen
  \bibfield  {author} {\bibinfo {author} {\bibfnamefont {S.-H.}\ \bibnamefont
  {Sun}}\ and\ \bibinfo {author} {\bibfnamefont {L.-M.}\ \bibnamefont
  {Liang}},\ }\bibfield  {title} {\bibinfo {title} {Experimental demonstration
  of an active phase randomization and monitor module for quantum key
  distribution},\ }\href {https://doi.org/10.1063/1.4746402} {\bibfield
  {journal} {\bibinfo  {journal} {Applied Physics Letters}\ }\textbf {\bibinfo
  {volume} {101}},\ \bibinfo {pages} {071107} (\bibinfo {year}
  {2012})}\BibitemShut {NoStop}%
\bibitem [{\citenamefont {Sibson}\ \emph
  {et~al.}(2017{\natexlab{b}})\citenamefont {Sibson}, \citenamefont {Erven},
  \citenamefont {Godfrey}, \citenamefont {Miki}, \citenamefont {Yamashita},
  \citenamefont {Fujiwara}, \citenamefont {Sasaki}, \citenamefont {Terai},
  \citenamefont {Tanner}, \citenamefont {Natarajan} \emph
  {et~al.}}]{Sibson2017}%
  \BibitemOpen
  \bibfield  {author} {\bibinfo {author} {\bibfnamefont {P.}~\bibnamefont
  {Sibson}}, \bibinfo {author} {\bibfnamefont {C.}~\bibnamefont {Erven}},
  \bibinfo {author} {\bibfnamefont {M.}~\bibnamefont {Godfrey}}, \bibinfo
  {author} {\bibfnamefont {S.}~\bibnamefont {Miki}}, \bibinfo {author}
  {\bibfnamefont {T.}~\bibnamefont {Yamashita}}, \bibinfo {author}
  {\bibfnamefont {M.}~\bibnamefont {Fujiwara}}, \bibinfo {author}
  {\bibfnamefont {M.}~\bibnamefont {Sasaki}}, \bibinfo {author} {\bibfnamefont
  {H.}~\bibnamefont {Terai}}, \bibinfo {author} {\bibfnamefont {M.~G.}\
  \bibnamefont {Tanner}}, \bibinfo {author} {\bibfnamefont {C.~M.}\
  \bibnamefont {Natarajan}}, \emph {et~al.},\ }\bibfield  {title} {\bibinfo
  {title} {Chip-based quantum key distribution},\ }\href
  {https://doi.org/10.1038/ncomms13984} {\bibfield  {journal} {\bibinfo
  {journal} {Nature Communications}\ }\textbf {\bibinfo {volume} {8}},\
  \bibinfo {pages} {13984} (\bibinfo {year} {2017}{\natexlab{b}})}\BibitemShut
  {NoStop}%
\bibitem [{\citenamefont {Bunandar}\ \emph
  {et~al.}(2018{\natexlab{b}})\citenamefont {Bunandar}, \citenamefont
  {Lentine}, \citenamefont {Lee}, \citenamefont {Cai}, \citenamefont {Long},
  \citenamefont {Boynton}, \citenamefont {Martinez}, \citenamefont {DeRose},
  \citenamefont {Chen}, \citenamefont {Grein} \emph
  {et~al.}}]{PhysRevX.8.021009}%
  \BibitemOpen
  \bibfield  {author} {\bibinfo {author} {\bibfnamefont {D.}~\bibnamefont
  {Bunandar}}, \bibinfo {author} {\bibfnamefont {A.}~\bibnamefont {Lentine}},
  \bibinfo {author} {\bibfnamefont {C.}~\bibnamefont {Lee}}, \bibinfo {author}
  {\bibfnamefont {H.}~\bibnamefont {Cai}}, \bibinfo {author} {\bibfnamefont
  {C.~M.}\ \bibnamefont {Long}}, \bibinfo {author} {\bibfnamefont
  {N.}~\bibnamefont {Boynton}}, \bibinfo {author} {\bibfnamefont
  {N.}~\bibnamefont {Martinez}}, \bibinfo {author} {\bibfnamefont
  {C.}~\bibnamefont {DeRose}}, \bibinfo {author} {\bibfnamefont
  {C.}~\bibnamefont {Chen}}, \bibinfo {author} {\bibfnamefont {M.}~\bibnamefont
  {Grein}}, \emph {et~al.},\ }\bibfield  {title} {\bibinfo {title}
  {Metropolitan quantum key distribution with silicon photonics},\ }\href
  {https://doi.org/10.1103/PhysRevX.8.021009} {\bibfield  {journal} {\bibinfo
  {journal} {Phys. Rev. X}\ }\textbf {\bibinfo {volume} {8}},\ \bibinfo {pages}
  {021009} (\bibinfo {year} {2018}{\natexlab{b}})}\BibitemShut {NoStop}%
\bibitem [{\citenamefont {Cao}\ \emph {et~al.}(2015)\citenamefont {Cao},
  \citenamefont {Zhang}, \citenamefont {Lo},\ and\ \citenamefont {Ma}}]{Lo_Ma}%
  \BibitemOpen
  \bibfield  {author} {\bibinfo {author} {\bibfnamefont {Z.}~\bibnamefont
  {Cao}}, \bibinfo {author} {\bibfnamefont {Z.}~\bibnamefont {Zhang}}, \bibinfo
  {author} {\bibfnamefont {H.-K.}\ \bibnamefont {Lo}},\ and\ \bibinfo {author}
  {\bibfnamefont {X.}~\bibnamefont {Ma}},\ }\bibfield  {title} {\bibinfo
  {title} {Discrete-phase-randomized coherent state source and its application
  in quantum key distribution},\ }\href
  {https://doi.org/10.1088/1367-2630/17/5/053014} {\bibfield  {journal}
  {\bibinfo  {journal} {New Journal of Physics}\ }\textbf {\bibinfo {volume}
  {17}},\ \bibinfo {pages} {053014} (\bibinfo {year} {2015})}\BibitemShut
  {NoStop}%
\bibitem [{\citenamefont {Currás-Lorenzo}\ \emph {et~al.}(2021)\citenamefont
  {Currás-Lorenzo}, \citenamefont {Wooltorton},\ and\ \citenamefont
  {Razavi}}]{Guillermo_2021}%
  \BibitemOpen
  \bibfield  {author} {\bibinfo {author} {\bibfnamefont {G.}~\bibnamefont
  {Currás-Lorenzo}}, \bibinfo {author} {\bibfnamefont {L.}~\bibnamefont
  {Wooltorton}},\ and\ \bibinfo {author} {\bibfnamefont {M.}~\bibnamefont
  {Razavi}},\ }\bibfield  {title} {\bibinfo {title} {Twin-field quantum key
  distribution with fully discrete phase randomization},\ }\href
  {https://doi.org/10.1103/PhysRevApplied.15.014016} {\bibfield  {journal}
  {\bibinfo  {journal} {Phys. Rev. Applied}\ }\textbf {\bibinfo {volume}
  {15}},\ \bibinfo {pages} {014016} (\bibinfo {year} {2021})}\BibitemShut
  {NoStop}%
\bibitem [{\citenamefont {Renner}\ and\ \citenamefont {Cirac}(2009)}]{finetti}%
  \BibitemOpen
  \bibfield  {author} {\bibinfo {author} {\bibfnamefont {R.}~\bibnamefont
  {Renner}}\ and\ \bibinfo {author} {\bibfnamefont {J.~I.}\ \bibnamefont
  {Cirac}},\ }\bibfield  {title} {\bibinfo {title} {de finetti representation
  theorem for infinite-dimensional quantum systems and applications to quantum
  cryptography},\ }\href {https://doi.org/10.1103/PhysRevLett.102.110504}
  {\bibfield  {journal} {\bibinfo  {journal} {Phys. Rev. Lett.}\ }\textbf
  {\bibinfo {volume} {102}},\ \bibinfo {pages} {110504} (\bibinfo {year}
  {2009})}\BibitemShut {NoStop}%
\bibitem [{\citenamefont {Currás-Lorenzo}\ \emph {et~al.}(2022)\citenamefont
  {Currás-Lorenzo}, \citenamefont {Tamaki},\ and\ \citenamefont
  {Curty}}]{Guillermo}%
  \BibitemOpen
  \bibfield  {author} {\bibinfo {author} {\bibfnamefont {G.}~\bibnamefont
  {Currás-Lorenzo}}, \bibinfo {author} {\bibfnamefont {K.}~\bibnamefont
  {Tamaki}},\ and\ \bibinfo {author} {\bibfnamefont {M.}~\bibnamefont
  {Curty}},\ }\bibfield  {title} {\bibinfo {title} {Security of decoy-state
  quantum key distribution with imperfect phase randomization},\ }\href
  {https://arxiv.org/abs/2210.08183} {\bibfield  {journal} {\bibinfo  {journal}
  {preprint arXiv:2210.08183}\ } (\bibinfo {year} {2022})}\BibitemShut
  {NoStop}%
\bibitem [{\citenamefont {Lo}(2005)}]{vacuum}%
  \BibitemOpen
  \bibfield  {author} {\bibinfo {author} {\bibfnamefont {H.-K.}\ \bibnamefont
  {Lo}},\ }\bibfield  {title} {\bibinfo {title} {Getting something out of
  nothing},\ }\href {https://doi.org/10.26421/QIC5.45-10} {\bibfield  {journal}
  {\bibinfo  {journal} {Quantum Inf. Comput.}\ }\textbf {\bibinfo {volume}
  {5}},\ \bibinfo {pages} {413} (\bibinfo {year} {2005})}\BibitemShut {NoStop}%
\bibitem [{\citenamefont {Gottesman}\ \emph {et~al.}(2004)\citenamefont
  {Gottesman}, \citenamefont {Lo}, \citenamefont {L\"utkenhaus},\ and\
  \citenamefont {Preskill}}]{GLLP}%
  \BibitemOpen
  \bibfield  {author} {\bibinfo {author} {\bibfnamefont {D.}~\bibnamefont
  {Gottesman}}, \bibinfo {author} {\bibfnamefont {H.-K.}\ \bibnamefont {Lo}},
  \bibinfo {author} {\bibfnamefont {N.}~\bibnamefont {L\"utkenhaus}},\ and\
  \bibinfo {author} {\bibfnamefont {J.}~\bibnamefont {Preskill}},\ }\bibfield
  {title} {\bibinfo {title} {Security of quantum key distribution with
  imperfect devices},\ }\href {https://doi.org/10.26421/QIC4.5-1} {\bibfield
  {journal} {\bibinfo  {journal} {Quantum Information and Computation}\
  }\textbf {\bibinfo {volume} {4}},\ \bibinfo {pages} {325} (\bibinfo {year}
  {2004})}\BibitemShut {NoStop}%
\bibitem [{\citenamefont {Koashi}(2009)}]{Koashi2009}%
  \BibitemOpen
  \bibfield  {author} {\bibinfo {author} {\bibfnamefont {M.}~\bibnamefont
  {Koashi}},\ }\bibfield  {title} {\bibinfo {title} {Simple security proof of
  quantum key distribution based on complementarity},\ }\href
  {https://doi.org/10.1088/1367-2630/11/4/045018} {\bibfield  {journal}
  {\bibinfo  {journal} {New J. Phys.}\ }\textbf {\bibinfo {volume} {8}},\
  \bibinfo {pages} {045018} (\bibinfo {year} {2009})}\BibitemShut {NoStop}%
\bibitem [{\citenamefont {Tamaki}\ \emph {et~al.}(2014)\citenamefont {Tamaki},
  \citenamefont {Curty}, \citenamefont {Kato}, \citenamefont {Lo},\ and\
  \citenamefont {Azuma}}]{Tamaki}%
  \BibitemOpen
  \bibfield  {author} {\bibinfo {author} {\bibfnamefont {K.}~\bibnamefont
  {Tamaki}}, \bibinfo {author} {\bibfnamefont {M.}~\bibnamefont {Curty}},
  \bibinfo {author} {\bibfnamefont {G.}~\bibnamefont {Kato}}, \bibinfo {author}
  {\bibfnamefont {H.-K.}\ \bibnamefont {Lo}},\ and\ \bibinfo {author}
  {\bibfnamefont {K.}~\bibnamefont {Azuma}},\ }\bibfield  {title} {\bibinfo
  {title} {Loss-tolerant quantum cryptography with imperfect sources},\ }\href
  {https://doi.org/10.1103/PhysRevA.90.052314} {\bibfield  {journal} {\bibinfo
  {journal} {Physical Review A}\ }\textbf {\bibinfo {volume} {90}},\ \bibinfo
  {pages} {052314} (\bibinfo {year} {2014})}\BibitemShut {NoStop}%
\bibitem [{\citenamefont {Nahar}(2022)}]{Nahar}%
  \BibitemOpen
  \bibfield  {author} {\bibinfo {author} {\bibfnamefont {S.}~\bibnamefont
  {Nahar}},\ }\emph {\bibinfo {title} {Decoy-State Quantum Key Distribution
  with Arbitrary Phase Mixtures and Phase Correlations}},\ \href@noop {}
  {Master's thesis},\ \bibinfo  {school} {University of Waterloo} (\bibinfo
  {year} {2022})\BibitemShut {NoStop}%
\bibitem [{\citenamefont {Upadhyaya}\ \emph {et~al.}(2021)\citenamefont
  {Upadhyaya}, \citenamefont {Himbeeck}, \citenamefont {Lin},\ and\
  \citenamefont {Lütkenhaus}}]{proj}%
  \BibitemOpen
  \bibfield  {author} {\bibinfo {author} {\bibfnamefont {T.}~\bibnamefont
  {Upadhyaya}}, \bibinfo {author} {\bibfnamefont {T.}~\bibnamefont {Himbeeck}},
  \bibinfo {author} {\bibfnamefont {J.}~\bibnamefont {Lin}},\ and\ \bibinfo
  {author} {\bibfnamefont {N.}~\bibnamefont {Lütkenhaus}},\ }\bibfield
  {title} {\bibinfo {title} {Dimension reduction in quantum key distribution
  for continuous- and discrete-variable protocols},\ }\href
  {https://doi.org/10.1103/PRXQuantum.2.020325} {\bibfield  {journal} {\bibinfo
   {journal} {PRX Quantum}\ }\textbf {\bibinfo {volume} {2}},\ \bibinfo {pages}
  {020325} (\bibinfo {year} {2021})}\BibitemShut {NoStop}%
\bibitem [{\citenamefont {Lo}\ and\ \citenamefont
  {Preskill}(2007)}]{Lo_Preskill}%
  \BibitemOpen
  \bibfield  {author} {\bibinfo {author} {\bibfnamefont {H.-K.}\ \bibnamefont
  {Lo}}\ and\ \bibinfo {author} {\bibfnamefont {J.}~\bibnamefont {Preskill}},\
  }\bibfield  {title} {\bibinfo {title} {Security of quantum key distribution
  using weak coherent states with nonrandom phases},\ }\href
  {https://doi.org/10.26421/QIC7.5-6-2} {\bibfield  {journal} {\bibinfo
  {journal} {Quantum Information and Computation}\ }\textbf {\bibinfo {volume}
  {8}},\ \bibinfo {pages} {431} (\bibinfo {year} {2007})}\BibitemShut {NoStop}%
\bibitem [{\citenamefont {Pereira}\ \emph {et~al.}(2020)\citenamefont
  {Pereira}, \citenamefont {Kate}, \citenamefont {Mizutani}, \citenamefont
  {Curty},\ and\ \citenamefont {Tamaki}}]{Pereira2020}%
  \BibitemOpen
  \bibfield  {author} {\bibinfo {author} {\bibfnamefont {M.}~\bibnamefont
  {Pereira}}, \bibinfo {author} {\bibfnamefont {G.}~\bibnamefont {Kate}},
  \bibinfo {author} {\bibfnamefont {A.}~\bibnamefont {Mizutani}}, \bibinfo
  {author} {\bibfnamefont {M.}~\bibnamefont {Curty}},\ and\ \bibinfo {author}
  {\bibfnamefont {K.}~\bibnamefont {Tamaki}},\ }\bibfield  {title} {\bibinfo
  {title} {Quantum key distribution with correlated sources},\ }\href
  {https://doi.org/10.1126/sciadv.aaz4487} {\bibfield  {journal} {\bibinfo
  {journal} {Science Advances}\ }\textbf {\bibinfo {volume} {6}},\ \bibinfo
  {pages} {eaaz4487} (\bibinfo {year} {2020})}\BibitemShut {NoStop}%
\bibitem [{\citenamefont {Winter}(1999)}]{Winter}%
  \BibitemOpen
  \bibfield  {author} {\bibinfo {author} {\bibfnamefont {A.}~\bibnamefont
  {Winter}},\ }\bibfield  {title} {\bibinfo {title} {Coding theorem and strong
  converse for quantum channels},\ }\href {https://doi.org/10.1109/18.796385}
  {\bibfield  {journal} {\bibinfo  {journal} {IEEE Transactions on Information
  Theory}\ }\textbf {\bibinfo {volume} {45}},\ \bibinfo {pages} {2481}
  (\bibinfo {year} {1999})}\BibitemShut {NoStop}%
\bibitem [{\citenamefont {Farenick}\ and\ \citenamefont
  {Rahaman}(2017)}]{bures}%
  \BibitemOpen
  \bibfield  {author} {\bibinfo {author} {\bibfnamefont {D.}~\bibnamefont
  {Farenick}}\ and\ \bibinfo {author} {\bibfnamefont {M.}~\bibnamefont
  {Rahaman}},\ }\bibfield  {title} {\bibinfo {title} {Bures contractive
  channels on operator algebras},\ }\href
  {https://nyjm.albany.edu/j/2017/23-63p.pdf} {\bibfield  {journal} {\bibinfo
  {journal} {New York Journal of Mathematics}\ }\textbf {\bibinfo {volume}
  {23}},\ \bibinfo {pages} {1369–1393} (\bibinfo {year} {2017})}\BibitemShut
  {NoStop}%
\end{thebibliography}%

\end{document}